\pdfoutput=1
\documentclass{cta-author}

{}
{}
{}
\usepackage[normalem]{ulem}
\usepackage{cancel}
\usepackage{lipsum}
\usepackage{fancyhdr}

\begin{document}

\title{Network Traffic Control for Multi-homed End-hosts via SDN}

\author{\au{Anees Al-Najjar$^{1, \corr}$}, \au{Furqan Hameed Khan$^{2}$}, \au{Marius Portmann$^{2}$}}

\address{$^1${Oak Ridge National Laboratory, Oak Ridge, TN, USA}\\ 
$^2${School of ITEE, The University of Queensland, St Lucia,
Brisbane, QLD 4072, Australia}\\ 
\email{alnajjaram@ornl.gov}}

\begin{abstract}
Software Defined Networking (SDN) is an emerging technology of efficiently controlling and managing computer networks, such as in data centres, Wide Area Networks (WANs), as well as in ubiquitous communication.
In this paper, we explore the idea of embedding the SDN components, represented by SDN controller and virtual switch, in end-hosts to improve network performance. In particular,  we consider load balancing across multiple network interfaces on end-hosts with different link capacity scenarios. We have explored and implemented different SDN-based load balancing approaches based on OpenFlow software switches, and have demonstrated the feasibility and the potential of this approach. The proposed system has been evaluated with multipath transmission control protocol (MPTCP). Our results demonstrated the potential of applying the SDN concepts on multi-homed devices resulting in an increase in achieved throughput of 55\% compared to the legacy single network approach and 10\% compared to the MPTCP.
\end{abstract}

\maketitle

\section{Introduction}
\label{sec:introdcution}

Software Defined Networking (SDN) is a relatively new paradigm for controlling and managing computer networks. The new feature behind SDN is to separate and centralise the functionality of forwarding packets in traditional networking components from the actual packet forwarding mechanism. The logically centralised (software) \emph{controller} or \emph{Network Operating System} (NOS), provides an abstraction of the distributed nature of the forwarding elements (switches, routers) to higher layer networking applications, such as routing, traffic engineering etc. 

\begin{figure}[t]
  \centering
   \includegraphics [width=3.35in]{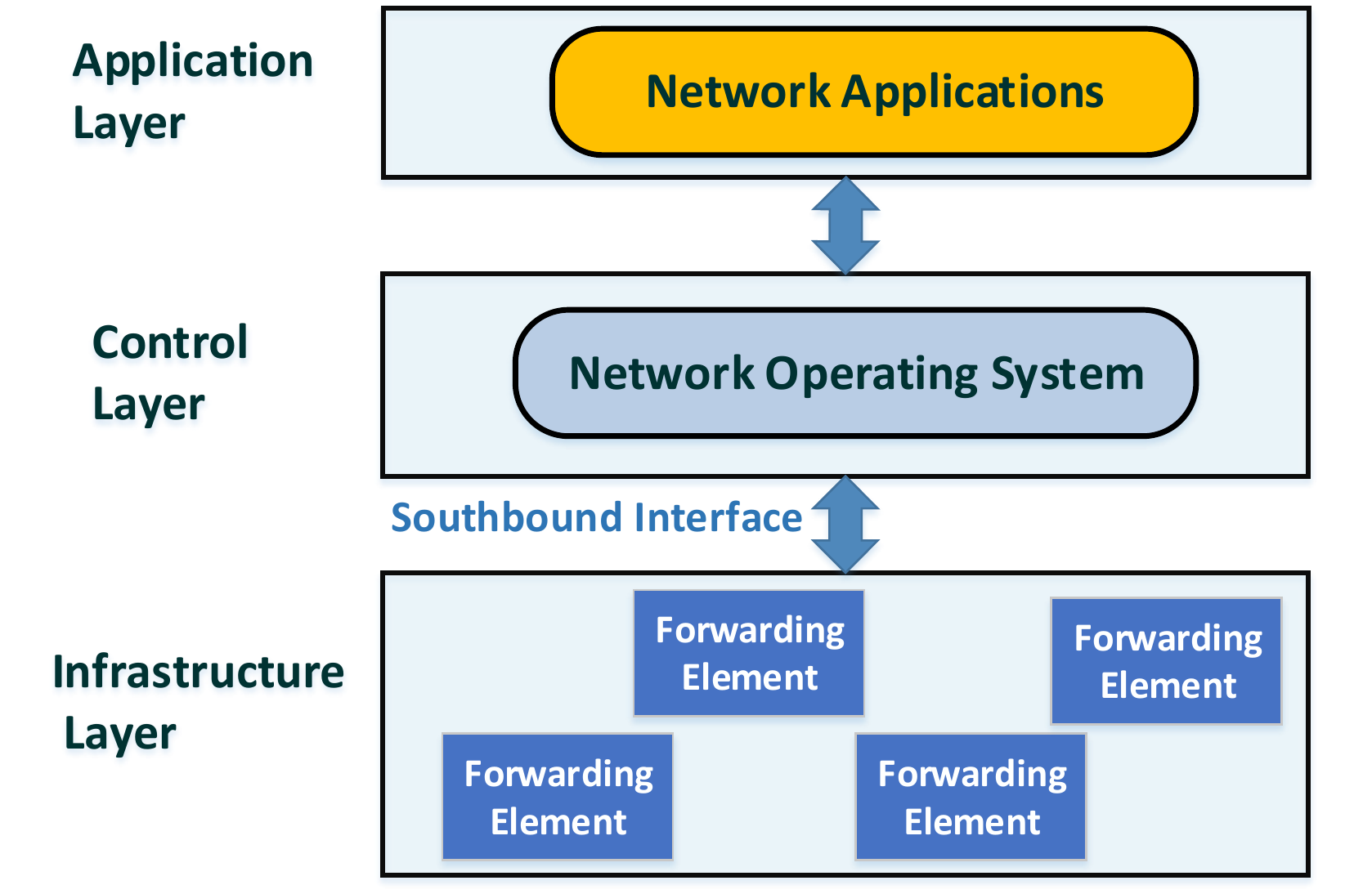}
    \caption{SDN Architecture}
  \label{fig:SDN-Architecture}
\end{figure}

Figure~\ref{fig:SDN-Architecture} shows the traditional SDN architecture~\cite{citeulike:12475417}. 
The bottom (\emph{infrastructure}) layer consists of a set of connected forwarding elements, i.e. SDN switches, which represent the data plane and provide basic packet forwarding functionality. In this paper, we extend this traditional view by adding end hosts to this layer.

The middle layer is the \emph{control layer} consists of a centralised SDN controller which implements the functionality of a NOS~\cite{GudeNOX2008}.
The NOS deals with and hides the distributed nature of the physical network, 
and provides an abstract view of network graph to higher layer services running over the \emph{application layer} of the SDN architecture~\cite{shenker2011future}. 
The SDN controller configures SDN switches by installing forwarding rules via the so called \emph{southbound interface}.
The predominant standard for this is OpenFlow \cite{mckeown2008openflow,urlOpenFlow2014} that will be discussed in more detail in Section~\ref{sec:Background}.

At the top of the SDN architecture is the application layer, where network applications and services, such as traffic engineering, routing, load balancing, etc. are implemented.
The research on SDN has so far mostly focused on the management of network infrastructure devices, i.e. forwarding elements, and has been highly successful in practical deployments, in particular in data centres and WANs~\cite{Hedera,B4}. 

In contrast to the existing body of research, in this paper we are exploring the idea of pushing SDN onto the end-host, which can be a normal computer, smartphone, a networked embedded device or a \emph{thing}.

For this, we consider the use case of network load balancing. We assume an end host with multiple network interfaces, e.g. a smartphone with 4G and WiFi, and consider the problem of efficiently balancing the user traffic across the available interfaces. When designing our mechanism, we attempted to meet the following objectives:
\begin{enumerate}
  \item be \textbf{efficient}, and not introduce a significant amount of overhead on the host.
  \item be \textbf{transparent} to both the applications as well as the rest of the network. Neither the applications nor the protocol stack of any other nodes in the network should have to be modified.
  \item \textbf{avoid packet reordering}, and the associated detrimental impact on TCP performance.
  \item work with any \textbf{OpenFlow compliant} software switch and controller, from version 1.3 onwards.
\end{enumerate}
Moreover, although most earlier research efforts on SDN-based network evaluations~\cite{SDNTrafficLBUsingMininet} use Mininet. Nonetheless, the Mininet network emulator has issues such as, it mainly supports wired network topologies, it does not offer greater network scalability and uses a shard Linux kernel space for all virtual hosts\cite{miminetlimitations}. The latter is greatly impact the design of the proposed system and having a separate control on the SDN based end-host from other network elements. Therefore, it is necessary to either enhance Mininet capability or use other evaluation tools that provide greater flexibility in one or the other domain. For this reason, a compatible network emulation testbed, Graphical Network Simulation (GNS3)\cite{gns3}, has been used in this work where the network elements are added as VMs with the support of separate kernel stack for each end-host VM.

While the idea of controlling network traffic in multi-homed end-hosts using SDN has been primarily explored in~\cite{al2016pushing}, in this paper, we extended that contribution via employing different load balancing algorithms with the aim of achieving the above goals. In~\cite{al2016pushing}, the evaluation was mainly conducted when the capacity of the links is static, while in this work the realistic network scenario with dynamic link capacity is considered. The algorithms are also evaluated with MultiPath TCP (MPTCP) \cite{rfc6897}. While the MPTCP protocol requires control over both ends of the connection, in contrast, the approach in this paper focuses on controlling only the end host connection. Our findings reveal that the SDN based traffic control over end-hosts can noticeably achieve higher performance compared with the use of single network interface. Further to that, our system demonstrated even better than MPTCP with the advantage of controlling the client side only.

The rest of the paper is organised as follows. Section \ref{sec:Background} briefly introduces OpenFlow and MPTCP. Section \ref{sec:Related Works} describes related works in applying SDN to end-host traffic control. Section \ref{sec:proposed system} explains our proposed approach and Section \ref{sec:loadBalancingApproaches} describes different load balancing algorithms used to control and distribute the application layer traffic on the end-host. Section \ref{sec:System Implementation} describes the system implementation and the experimental setup. Section \ref{sec:Evaluation for Static Link Capacity} and Section \ref{sec:Evaluation for Dynamic Link Capacity} demonstrate the evaluation results of the implemented system with static link and dynamic link capacity scenarios respectively. Finally, Section \ref{sec:Conclusion} summarises our findings and concludes the paper.
\section{Background}
\label{sec:Background}
This section provides an overview of a well-known south-bound interface (e.g. OpenFlow). Then we introduce a well-known transport layer technology (MPTCP) that enables utilising multiple network interfaces on the end-host for a single connection. 
\subsection{Openflow}
\label{subsec:OpenFlow}
OpenFlow~\cite{urlOpenFlow2014} is currently the dominant southbound interface protocol for SDN, which allows controllers to configure the forwarding behaviour of switches. It provides the interface between the infrastructure layer and the control layer, as shown in Figure~\ref{fig:SDN-Architecture}. OpenFlow allows a controller to install rules as \emph{flow table entries} in SDN switches via \emph{Flow-Mod} messages. The installed entries follow a \emph{match-action} paradigm and enable fine grained control over how packets are being forwarded in the network. 
Each packet arriving at a switch is \emph{matched} against the match fields of these rules, and the corresponding \emph{action/action list} of matching rule is executed. The supported match fields include packet header fields, such as IP source and destination address, MAC source and destination address, VLAN tags, etc. An action may ask a switch to perform one or more of the following,
\begin{itemize}
  \item Forward the arriving packet to controller or to next connected element by forwarding it through the corresponding port (\textit{action:output}).
  \item Modify/rewrite the specific packet header fields (\textit{action:set-field}). For example, this allows rewriting of IP and MAC addresses to implement address translation.
  \item Drop the packet (\textit{action:drop}).
\end{itemize}

The OpenFlow protocol allows gathering switch statistics that may leverage for achieving various network functionalities, like load balancing. Certain types of OpenFlow messages, such as \textit{Port Stats} and \textit{flow Stats} messages, can probe information related to links, ports and flow entries on the enabled switch(es). The controller can request statistics of active ports by sending a  \textit{PortStatsRequest} message to the switch; then, the latter replies with a \textit{PortStatsReply} message, with the probed information related to each port. Example of these are the cumulative number of sent and received packets and bytes, as well as the number of packets that have been dropped or had errors. The controller can also collect statistics about the active flow entries installed on the switch(es) through sending a \textit{FlowStatsRequest} message, upon which the switch replies with a \textit{FlowStatsReply} message. The message carries information related to each installed entry, for instance table\_id, priority,  number of bytes/packets that matched the rule, the active duration of the flow, and the match/action fields.

In this work, we will use the \emph{set-field} and  \emph{output} action features in our load balancing mechanisms discussed in Section~\ref{sec:loadBalancingApproaches} as well as \emph{port stats} message in the result validation in Section~\ref{sec:System Implementation}.


\subsection{MPTCP}
\label{subsec:MPTCP}

MPTCP is an extension of TCP that exploits multiple paths to deliver versatile network services to the end hosts. The use of MPTCP increases with the advent of new devices supporting operation with multiple networks (e.g. 3G/4G, WiFi). As a consequence, it is important to make efficient use of multiple paths at the same time. To address this issue,~\cite{bonaventure2012overview} shows that with careful monitoring and path updates it is possible to linearly increase network throughput multiple times for an MPTCP connection as new sub flows are added.

\begin{figure}[t]
  \centering
   \includegraphics [width=3.35in]{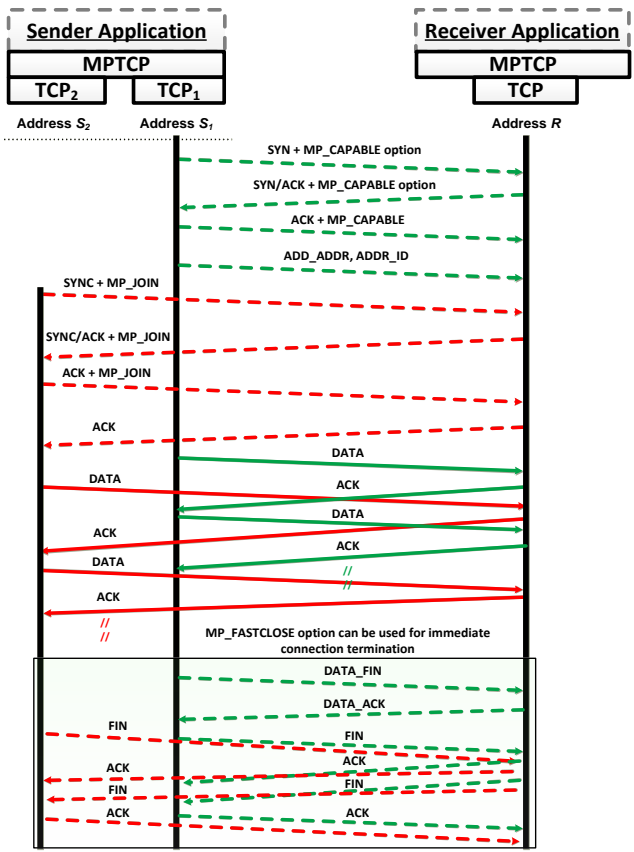}
    \caption{MPTCP Operations}
  \label{fig:mptcp-Design}
\end{figure}

As shown in Fig.~\ref{fig:mptcp-Design}, a multi-path (MPTCP) connection is basically a thin layer that operates in between the application and TCP layers and provides features for the creation, management, and termination of TCP sub-flows. Similar to the normal TCP operation, a sub-flow starts when the host exchange SYN, SYN-ACK, and ACK messages as shown in Fig.~\ref{fig:mptcp-Design}~\cite{ford2016tcp}\cite{mptcp}. The first SYN (handshake) message includes a new option called MP\_CAPABLE (in Fig.~\ref{fig:mptcp-Design}) to check if the other end supports MPTCP protocol; and if not, the connection falls to a normal TCP connection. In case the MPTCP is supported, a 64-bit authentication key as well as additional flags related to each end will be generated and exchanged via the added option. These fields are required in the later stages for the creation and authentication of TCP sub flows. A typical flow can be added/removed from the MPTCP connection using ADD\_ADDR/REMOVE\_ADDR fields added as an extension for MPTCP. Ones all data is sent, the sender use DATA\_FIN option to indicate that all data has been sent/acknowledged as shown in Fig.~\ref{fig:mptcp-Design}. The sender then closes the MPTCP connection over each sub-flow using the FIN flag (as in TCP) which is later acknowledged by the other end. To achieve backward compatibility, the normal RST/FIN signal are exchanged for each sub-flow in MPTCP, as shown in the connection termination phase in Figure~\ref{fig:mptcp-Design}. To terminate an ongoing session immediately, an MPTCP level closure (called MP\_FASTCLOSE) is also included. This command tells the peer entity that the connection needs to be immediately closed and no data will be accepted anymore~\cite{ford2016tcp}. Upon reception, the other end either sends the MP\_FASTCLOSE to close the whole connection or it can send the TCP RST on each sub-flows to close them one by one.

In a new transport layer solution such as MPTCP it is challenging to devise flow control and congestion control approaches that achieves optimal network throughput. In other words, flows in an MPTCP connection should achieve their throughput without unfairly affecting the other (normal TCP) flows in the network. Moreover, the deployment of MPTCP needs control over the operation of both ends during the connection life-time. 

Traditional Linux kernels (2.6.19 onwards) and some Windows operating system editions use TCP cubic as a default congestion control algorithm~\cite{balasubramanian2017updates} which is optimised for high bandwidth and high latency networks. Compared with other congestion control algorithms, TCP cubic uses the time since most recent congestion event to update the congestion window. The size of congestion window follows a concave shape before the congestion event, after a congestion detected; the window size follows a convex shape. 
%

\section{Related Works}
\label{sec:Related Works}
SDN and OpenFlow have been used for load balancing in previous works, however, mostly at the server and in the network infrastructure. 

In previous works, the SDN based load balancing solutions are considered running over servers inside the network infrastructure. For example, the~\textit{Plug-n-Serve} solution presented in \cite{handigol2009plug} provides an SDN based solution for minimizing the HTTP request response time. The idea is to balance the web traffic (HTTP requests) across a number of web servers and network paths. As another example, the OpenFlow-based server load balancing is presented in~\cite{wang2011openflow}. The paper addresses the problem of scalability in Data Centre Networks (DCN) by using the wildcard flow rules. The load balancing decisions are made based on source IP addresses, which is not applicable in our scenario of client-side load balancing.

\par

An initial effort of developing network solutions over end-host is the Eden system~\cite{ballani2015enabling}. The work allows end-hosts to implement a wide range of application-aware networking functions, including path-based load balancing. However, this work does not address the issue of balancing traffic load across multiple host interfaces. Another key contrast to our work is that the Eden system is not transparent to applications or the network infrastructure, and hence cannot be easily deployed.

\par

In~\cite{mptcp}, the authors present a multipathing and load balancing solution based on OpenFlow. The proposed method uses OpenFlow only for the network infrastructure and not for the end-host. Furthermore, end-hosts (client and server) need to use MPTCP as the transport protocol to make use of the load balancing feature. Different previous works propose new congestion control algorithms for MPTCP~\cite{paasch2013multipath}. For example, alias Linked Increase Algorithm (LIA), alias Opportunistic Linked Increase Algorithm (OLIA), alias Delay-based Congestion Control for MPTCP wVegas, and alias Balanced Linked Adaptation Congestion Control Algorithm (BALIA). Likewise different approaches for sub-flows management has been proposed to efficiently distribute MPTCP traffic between different sub-flows. To reduce the complexity of multi-path problem in MPTCP, in this paper we suggest a client side solution that requires the control only one end. Similar to TCP, MPTCP also allocates traffic per packet basis. Contrary to this, in our work the traffic allocation is done on a per flow basis. In other words, our system runs over an end hosts and enable it to distribute traffic of each flow over the suitable network interface.

The authors of~\cite{yap2012making} present an approach that allows the use of multiple network interfaces on end-hosts. Their system is implemented in Android and uses an OpenFlow switch for controlling the network traffic. The paper discusses network handover, interface aggregation and dynamic interface selection. Most of the proposed mechanisms require both ends of the connection to support the special network protocol and stack, which is in contrast to our work. Furthermore, while discussing various aspects of using of multiple host interfaces, ~\cite{yap2012making} does not address the specific problem of load balancing.

Some initial efforts for implementing SDN-based load balancing solutions for end-devices are made in~\cite{8307045} and~\cite{8357765}. In~\cite{8307045}, the authors implement a proactive load balancing approach for SDN-based machine-to-machine (M2M) networks where short response time is desired. Real testbed evaluation of the proposed framework shows that it can minimize the device's response time to around 50\% of the non-SDN approaches. Note that in contrast to our work, both the OpenFlow capable hardware SDN-switch and the controller (running traffic-aware load balancing algorithm) used in their framework exist outside the end-device. Moreover, most earlier load balancing solutions in the SDN-based network environment uses Mininet~\cite{mininet}. For example, in~\cite{8357765} ~\cite{SDNTrafficLBUsingMininet} the work uses mininet to implement traffic load balancing scheme using the virtual SDN (vSDN) controller.~\cite{8357765} develop an approach of SDN based load balancing for underwater acoustic sensor networks (UASN) with multiple controllers. In the proposed proactive load balancing algorithm~\cite{SDNTrafficLBUsingMininet}, the primary vSDN controller creates a secondary virtual SDN controller as its new copy based on traffic load balancing demand. The overall process of new controller inclusion is transparent to the end-user. In this way, the control plane can be extended based on traffic load leading to about 50\% reduction in average load and 41\% decrease in the average delay. Other proposals of traffic load balancing include~\cite{SDNLoadBalancing}, where the authors implemented a customized OpenFlow based SDN controller to distribute traffic in the backhaul and access of cellular networks. In contrast to the above contributions, this paper aims to achieve traffic load balancing across multiple end-host network interfaces. This paper is an extension of our earlier work where SDN was applied for traffic control on multi-homed end devices \cite{al2016pushing}. This earlier work demonstrated the basic concept, but was limited to the case of static link capacities. In contrast, this paper considers the case of dynamic link capacity, which is critical in wireless networks. A key contribution of this paper is an extensive experimental evaluation based on a realistic wireless network scenario, as well as a performance comparison with MPTCP.

%

\section{ Proposed system}
\label{sec:proposed system}

\begin{figure}[t]
  \centering
   \includegraphics [width=3in]{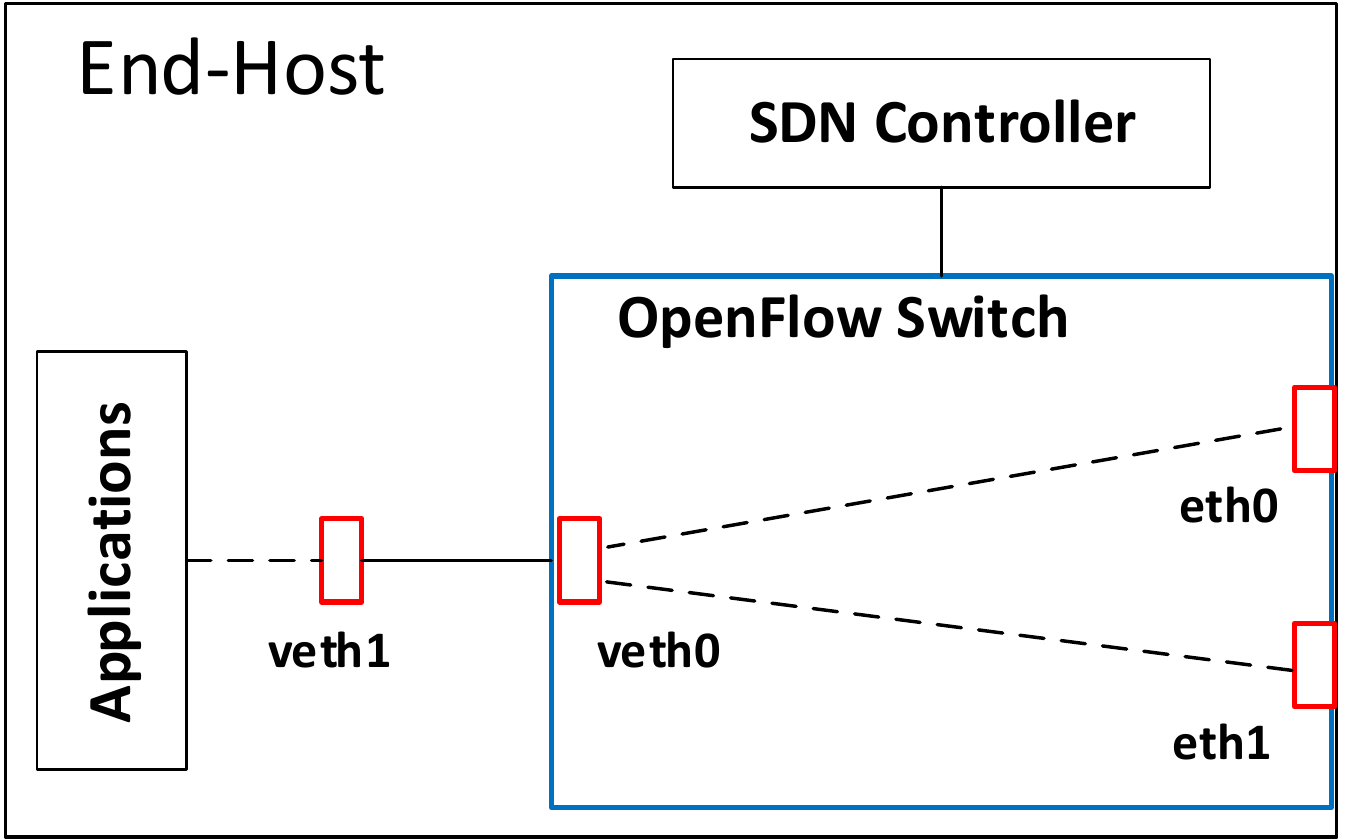}
    \caption{System Architecture}
  \label{fig:System-Architecture}
\end{figure}

The basic architecture of our host-based network load balancing mechanism using OpenFlow is shown in Figure~\ref{fig:System-Architecture}. In this scenario, the host has only two interfaces, but our approach is generalised for any number of interfaces.

\par
OpenFlow switch is the key component that facilitates the load balancing on the host with its ability to control the flow of network packets.
The switch is configured to control the host's external network interfaces, in this case \emph{eth0} and \emph{eth1} in Figure~\ref{fig:System-Architecture}. 
In order to provide the required level of indirection to switch traffic between these interfaces in a way that is transparent to the application, we also configured a pair of (connected) virtual network interfaces (shown in Figure~\ref{fig:System-Architecture} as \emph{veth0} and \emph{veth1}). Interface \emph{veth0} is attached to the switch and \emph{veth1} is configured as an \emph{internal gateway} via which all application traffic is sent. This is achieved via the configuration of a corresponding default route in the kernel's routing table.


With this setup, the switch can implement traffic load balancing by choosing an external interface e.g. \emph{eth0} or \emph{eth1} through which the packets are sent, depending on the match-action rules installed on the switch by the controller\footnote{In our system, the controller is co-located on the host. However, this is not a requirement, and in future work we will explore the idea of delegating
control to a remote controller.}.
 
 
 From the system architecture of Figure~\ref{fig:System-Architecture}, as we have detached the application from directly communicating via an external facing interface, we need to address a couple of issues, i.e. address translation and ARP handling.
Address translation is required, since packets leaving the host need to have the correct IP source address as well as the MAC source and destination address, depending on which interface was chosen as the egress interface by the load balancer. This needs to be implemented for packets in both the forward and reverse directions. OpenFlow provides a packet header rewriting mechanism, which allows the implementation of address translation for both IP and MAC addresses. This is achieved in OpenFlow by adding a corresponding \emph{set-field} action prior to the \emph{output} action.

The other issue is ARP handling. In a traditional system, when a host tries to send an IP packet to a particular destination, it would look up the address of the next hop node and the corresponding interface in the routing table. Then, an ARP request is issued in order to establish the MAC address that belongs to the next hop IP address. The ARP request is broadcasted on the local network, and the node with the specified IP address answers with an ARP reply message that contains its MAC address. Due to the level of indirection introduced in our setup, this does not work. We therefore implement a Proxy ARP mechanism, in which the switch intercepts any ARP requests from the host and sends them to the controller. 
The controller then instructs the switch to send an ARP reply message with the required MAC address.


\section{Load Balancing Approaches}
\label{sec:loadBalancingApproaches}
Since a key requirement of our load balancing approach is the avoidance of packet reordering in a TCP session, we need to guarantee that all packets belonging to the same TCP session are sent via the same host network interface. To achieve this, we perform load balancing at the level of granularity of TCP connections \footnote{While our discussions in the paper is focused on TCP traffic, the same basic load balancing approach can be applied to UDP traffic~\cite{al2018flow}.}.
We consider two basic approaches of SDN-based load balancing on the end host, to which we refer as the \emph{controller-based} and \emph{switch-based} approaches.
These methods are discussed in the following.

\subsection{Controller-Based load balancing approach}
\label{subsec:Controller-Based load balancing approach}
In this approach, load balancing decisions for each individual TCP session are made by the SDN controller. The controller is responsible for allocating a TCP flow for a specific network interface. The switch sends the first packet of each TCP session, referred as TCP SYN packet, which is initiated by an end-host application to the controller via Packet\_In message. Among multiple interfaces and based upon the load balancing algorithm, the controller will decide which interface should be chosen for forwarding that TCP flow. Then, the controller will install a flow entry on the switch via Flow-Mod message to send out not just the first packet, but also the rest in that TCP session via the selected interface.

 Algorithm~\ref{alg:algorithm1} shows the processing of packets received from the switch at the controller. The controller checks if the packet is a TCP SYN packet, and then assigns an interface $i$ to the new flow, from the available $N$ interfaces using a certain interface selecting algorithm \textit{Link\_Select\_Algo()} discussed later in this section. Once the interface is chosen, the controller initiates an OpenFlow rule $R$, with 3 match fields (lines 6-8) and 4 actions (lines 9-12). The rule will match on TCP/IP packets and the specific source port of the received packet (line 8). In our scenario, the TCP source port will uniquely identify all TCP packets belonging to this session.
The \emph{setField} actions in lines 9-11 provide the required address translation, as discussed earlier.
The \emph{output} action in line 12 will instruct the switch to forward the packet via the chosen interface $i$.
Finally, line 13 installs the rule $R$ on the switch.

At this point, the switch sent all packets belonging to this TCP session via the chosen interface, without any further involvement of the controller.

The load balancing algorithms are categorised into \textit{stateless} and \textit{stateful} algorithms. While the former category does not take the link status parameters (e.g. bandwidth, the level of congestion, delay characteristics, etc.) into consideration when opting the link, the latter considers link parameters in the link selection process.


\begin{algorithm}
\caption{Controller-based Load Balancing}
\label{alg:algorithm1}
\begin{algorithmic}[1]
\STATE {$flowCounter \gets 0$}

\FOR{\textbf{each} Packet-In Event with $pkt$}
\IF {$pkt.flag.SYN==1$ \AND $pkt.flag.ACK==0$}

\STATE   {$i \gets Link\_Select\_Algo()$}
\STATE{$flowCounter \gets flowCounter+1$}

\STATE{$R$.match[0]$\gets eth\_type== IP$}
\STATE{$R$.match[1]$\gets ip\_proto== TCP$}
\STATE{$R$.match[2]$\gets tcp\_src\_port== pkt.tcp.src\_port$}

\STATE{$R$.action[0]$\gets setField(ipv4\_src=IP\_addr[i])$}
\STATE{$R$.action[1]$\gets setField(eth\_src=MAC\_addr[i])$}
\STATE{$R$.action[2]$\gets setField(eth\_dst=GW\_MAC[i])$}
\STATE{$R$.action[3]$\gets output(i)$}

\STATE{sendFlowModMessage($R$)} 
\ENDIF
\ENDFOR

\end{algorithmic}
\end{algorithm}


\subsubsection{Stateless Load Balancing: Round Robin (RR) Algorithm}
\label{subsubsec:Round Robin (RR) Algorithm}

In this category, we implement a simple approach that alternatively allocates the TCP flows among the available network interfaces without taking into account the link parameters.

The process of opting the network interface (step 4 in Algorithm \ref{alg:algorithm1}) is done by allocating the new flow to the next available network interface which is simply done via the modulo operation as stated below. 
 $$i \gets flowCounter \mod(N)$$
 \noindent Where:

 \textit{flowCounter} represents the increasing flow counter and $N$ is the number of available network interfaces connected to the end-host.


\subsubsection{Stateful Link Selection}
\label{ubsubsec:Stateful Link Selection}

 Contrary to stateless link selection, in this approach a link/interface is chosen based on its current state. The proposed system architecture is capable of controlling network traffic via utilizing different network parameters, such as delay, packet loss and available bandwidth. In a previous paper~\cite{al2018enhancing}, we have demonstrated how packet-loss ratio and delay can be used for selecting the optimal network interface for VoIP sessions. In contrast, this paper considers the available link bandwidth as the criterion for allocating flows to network interfaces.

 Two types of controller-based-stateful load balancing algorithms will be discussed in this part, namely: Maximum Bandwidth (MBW) and the Weighted Round Robin (WRR) link selection.
\begin{enumerate}[i)]
    \item Maximum Bandwidth (MBW) Link Selection:
    \label{subsubsecitem:Maximum Bandwidth (MBW) Link Selection}
In this algorithm, the controller allocates a TCP flow to the link having maximum bandwidth among the available links. Hence, the step 4 for selecting the interface in Algorithm~\ref{alg:algorithm1} is determined from the following equation.

\begin{equation}
i \gets \operatorname*{argmax}_{1\leq j\leq N}\{ BW_j (t_0) \} \label{eq:MAX_Int} 
\end{equation}\\
Where:
\begin{conditions*}
i & selected network interface\\
BW_j (t_0) & bandwidth of link $j$ at time $t_0$\\
N & number of available network interfaces
\end{conditions*}

The above equation utilises \textit{argmax} function that selects the interface $i$ that has the maximum bandwidth $BW$ among the available network interfaces $N$ for allocating a new flow. For computing the link bandwidth, a link monitoring module runs in the background that collects and computes link status parameters such as the bandwidth per time basis\cite{al2016link}. The bandwidth computation is briefly explained in section \ref{sec:Evaluation for Dynamic Link Capacity}.

As stated, our approach utilises one link (having the maximum residual bandwidth) among the available links. Note that this algorithm achieves uniform traffic distribution only over a small time duration. This is because it allocates new flows to the instantaneous maximum residual bandwidth link and continue forwarding the previously allocated flows to the link that had the maximum bandwidth. In the next algorithm, we focus on achieving a better link utilisation over the whole duration through load balancing.

\item{Weighted Round Robin (WRR) Algorithm}
\label{subsubsecitem:Weighted Round Robin (WRR) Algorithm}

 The previous algorithms either alternatively selects the network interface without considering its status (as the use of RR) or they choose the link with highest bandwidth (when MBW is used). Both methods lacks in efficient utilisation of available links for simultaneous traffic forwarding which can degrade the network performance. 
 
One of the approaches that try to improve the link utilisation and addresses the previously mentioned shortcomings is Weighted Round Robin (WRR). The approach proportionally allocates network flows based on the available bandwidth of the links as described below.

Firstly, we find the weights of the links based on their available bandwidth as stated below.

\begin{equation}
\label{eq02}
W_{L{_{j/t_0}}}=\tfrac{BW_{j} (t_0)}{\sum_{k=1}^{N} BW_{k} (t_0)}*100\ , j\in\left \{ 1:N \right \}
\end{equation}

\indent Where:
\begin{conditions*}
W_{L{_{j/t_0}}} & weight of link $j$ computed at time $t_0$\\
{BW_{j} (t_0)} & bandwidth of link $j$ at time $t_0$\\
N & number of available links/ interfaces on the end-host\\
\end{conditions*}

The weight of a certain link at time $t_0$ is the percentage of the bandwidth of that link over the total bandwidth of the available links at that time. Similar to the BW process computation, the process of computing the links' weight is also done as a background process.

To ensure fair bandwidth utilisation of available links, traffic allocation should be done in an interleaved manner. Otherwise, the system either ends up using only one interface or it may inefficiently select links to forward traffic.
Algorithm \ref{alg:WRR-algorithm} explains the \textit{Link\_Select\_Algo()} function that selects the network interface using WRR algorithm. Briefly, the algorithm proportionally selects a network interface based on its weight, after which the weight is decremented. So, the algorithm first makes sure that the weight dictionary $W$, which represents the tuples of available links and their weights $(N,w_N)$, is not empty during the time over which the bandwidth and the weights are measured. If the condition is false, the weights are recomputed as shown in Algorithm 2 (steps 2-4). Note that the function $Compute\_Weight$ ($N$, $BW$) uses Equation~\ref{eq02} to calculate weights of the corresponding link. Then, the interface is chosen similar to RR but based on links weights (step 5). This is followed by decreasing the weight of the chosen interface and then checking if the link weight becomes zero to be removed from the dictionary (steps 6-9). Thus, the link is chosen as an output of Algorithm \ref{alg:WRR-algorithm} is assigned in step 4 of Algorithm \ref{alg:algorithm1}.
\end{enumerate}

In the following section, we explore an alternative load balancing mechanism that is switch-based, with only minimal involvement of the controller.

\begin{algorithm}[t]
\caption{Load Balancing Algorithm: WRR}
\label{alg:WRR-algorithm}

\begin{algorithmic}[1]
\STATE {$Link\_Select\_Algo(flowCounter,W)$}
\IF {W==NULL}
\STATE{W$\gets$ Compute\_Weight(N, BW)}
\ENDIF
\STATE{$i \gets flowCounter \mod(W)$}
\STATE{W[$i]\gets$ W[$i]$-1}
\IF {W[$i$]==0}
\STATE{del (W[$i$])}
\ENDIF
\STATE {return $i$}
\end{algorithmic}
\end{algorithm}


\subsection{Switched-Based Load Balancing Approach}
\label{subsec:Switched-Based Load Balancing Approach}

An OpenFlow switch provides a very limited set of primitives, and we can not run arbitrary code as we can do on the controller.
Therefore, we need a different approach to achieve load balancing at the switch, with only minimal involvement of the controller.

Since version 1.3, OpenFlow supports the concept of \emph{groups} and \emph{group tables}, which provide abstractions for sets of ports and a level of indirection that allows the implementation of features such as multicasting, fast failover, etc. Each group consists of a set of \emph{buckets}, and each of those buckets contains a set of actions that includes the switch port (host interface) via which the packet is to be forwarded. For each packet arriving at a group, one or more buckets are selected and the corresponding actions are performed. OpenFlow supports a number of different group types, \emph{All} for multicast or flooding, \emph{Indirect} to implement simple indirection, \emph{Fast Failover}, which simply selects the first live port, and \emph{Select}, which is the one we are going to use. In the \emph{Select} group type, only a single bucket, and corresponding action set are chosen and executed. Possible selection algorithms implemented by OpenFlow switches include Round Robin and hash-based selection. Unfortunately, we cannot use the Round Robin selection method for our load balancing method, since the selection is done on a per-packet basis, rather than per TCP session. This would cause packets belonging to the same TCP session to be spread across different host interfaces, resulting in most likely in packet reordering. 

In the hash-based selection method, the switch computes a hash function over a tuple of packet information, e.g. the typical address and protocol 5-tuple, and the choice of the bucket is based on the value of the hash. For example, in a scenario where we only have two interfaces (and buckets), the least significant bit of the hash value can be used for the selection. The hash-based selection guarantees that all packets belonging to the same TCP session are forwarded via the same interface.

The switch hash-based mechanism also allows Weighted Group Table selection for the \textit{Select} group table type. Likewise, the WRR controller-based approach the, the OpenFlow switch (version 1.3 onwards) can proportionally allocate the traffic to among the defined buckets.

Compared to the RR and hash-based switch approaches, there is an important difference between the methods. The hash-based selection implemented at the switch is (pseudo) random and can achieve equal load sharing, however, it cannot implement Round Robin load balancing.
The other key difference between the two approaches is that in the switch-based load balancing, the controller is only involved in the initial one-off installation of the group and flow table. After that, the per TCP session traffic load balancing is handled independently by the switch.


\section{System Implementation}
\label{sec:System Implementation}

\subsection{Experimental Setup}
\label{subsec:Experimental Setup}

We have implemented a prototype of our host-based traffic load balancing mechanism using OpenFlow.
We used Linux (Xubuntu) with kernel version 3.13.0-24 as the operating system for the end-host.

The Ryu \cite{ryu} open-source SDN framework forms the basis for our controller. Unlike other SDN controllers such as Open Network Operating System (ONOS~\cite{onosarticle}) and floodlight~\cite{Floodlightarticle}, Ryu is a more sophisticated and a lightweight option which makes it an impeccable choice to run over the end host. We developed a load balancing component inside Ryu (using Python) which makes decisions about the network traffic sent by the application running over end host. The Ryu controller sends OpenFlow instructions to the SDN software switch existing inside the end-host device as discussed earlier in Fig.~\ref{fig:System-Architecture}.

Two different OpenFlow software switches are implemented for our prototype, namely the \emph{Open vSwitch} (OVS) and \emph{ofdatapath}. OVS is an open source virtual SDN switch supporting the OpenFlow protocol, and is widely supported and used. OVS is supported and distributed with the Linux kernel (since its version 2.6.32). In this implementation, OVS version 2.4 is used.
\emph{Ofdatapath} is a user-space \mbox{OpenFlow 1.3} compatible switch, based on the Stanford OpenFlow 1.0 reference switch implementation~\cite{StanfordOpenFlowreferenceswitch}.

To evaluate the SDN-enabled network environment (e.g. data centers and WANs), a standard choice is to use Mininet. It allows emulating topologies with one or more SDN controllers~\cite{8357765}, multilevel switches, and multiple end-hosts which can be run in a shared Linux kernel space and network stack. However, as described earlier Mininet does not enable a fully standalone kernel space for the end-host VM which is necessary for our approach (for emulating embedding SDN components inside multi-homed end-hosts to control the network traffic transparently to the applications, network infrastructure, and server-side). Therefore to achieve separate network stack for each network component in our experimental topologies (shown in Fig.~\ref{fig:Baseline-Topology} and Fig.~\ref{fig:Network-Topo}) GNS3\cite{gns3} is used. This software allows the creation of virtual network topology consisting of nodes (such as hosts, routers, etc.) running a full operating system and network stack, connected via the virtual links.

The Linux traffic control tool~\emph{tc} was used to emulate different link capacities.
The HTTP traffic is considered in our experiments and the apache benchmark (~\emph{ab}~\cite{ab}) tool is used for the generation of HTTP GET requests,
and~\emph{wbox}~\cite{wbox} was used as a web server. We also used MPTCP~\cite{paasch2013multipath} version 0.91 for the evaluation purpose.
Finally, all our experiments were run on a single Dell PC with a Windows 7 host system, a Core i7 3.6GHz CPU and 16GB of RAM.

\subsection{OpenFlow Switch Baseline Performance}
\label{subsec:OpenFlow Switch Baseline Performance}
As described above, we considered two OpenFlow software switches for our host-based load balancing mechanism, OVS and ofdatapath.
As an initial experiment, our goal was to measure and compare the efficiency of the two versions, and also provide a comparison to 
a legacy networking stack without any SDN processing, as a baseline.

\begin{figure}[t]
  \centering
 \includegraphics [width=3.35in]{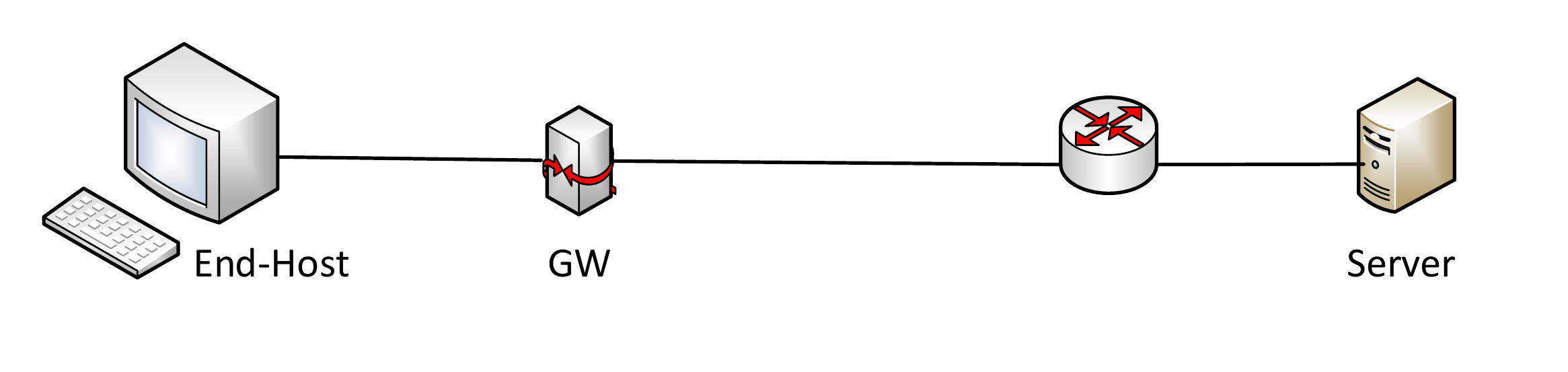}
  \caption{Scenario for Switch Performance Evaluation}
  \label{fig:Baseline-Topology}
\end{figure}

For this experiment, we used the scenario shown in Figure~\ref{fig:Baseline-Topology}, with a host that has a single interface connected to a server via a gateway and another IP router.
No capacity limit was imposed on the virtual links.

We considered OVS and ofdatapath as the OpenFlow switch in the configuration shown in Figure~\ref{fig:System-Architecture}.
However, in this case no load balancing scheme is run and we only use a single host interface, i.e. traffic is simply bridged between \emph{veth0} and \emph{eth0} via the installation of 
a corresponding rule in the switch. 
As a reference, we also considered the \emph{No SDN} case, in which the host has a traditional network stack configuration, with no SDN processing.

\begin{figure}[t]
  \centering
  \includegraphics[width=3.35in]{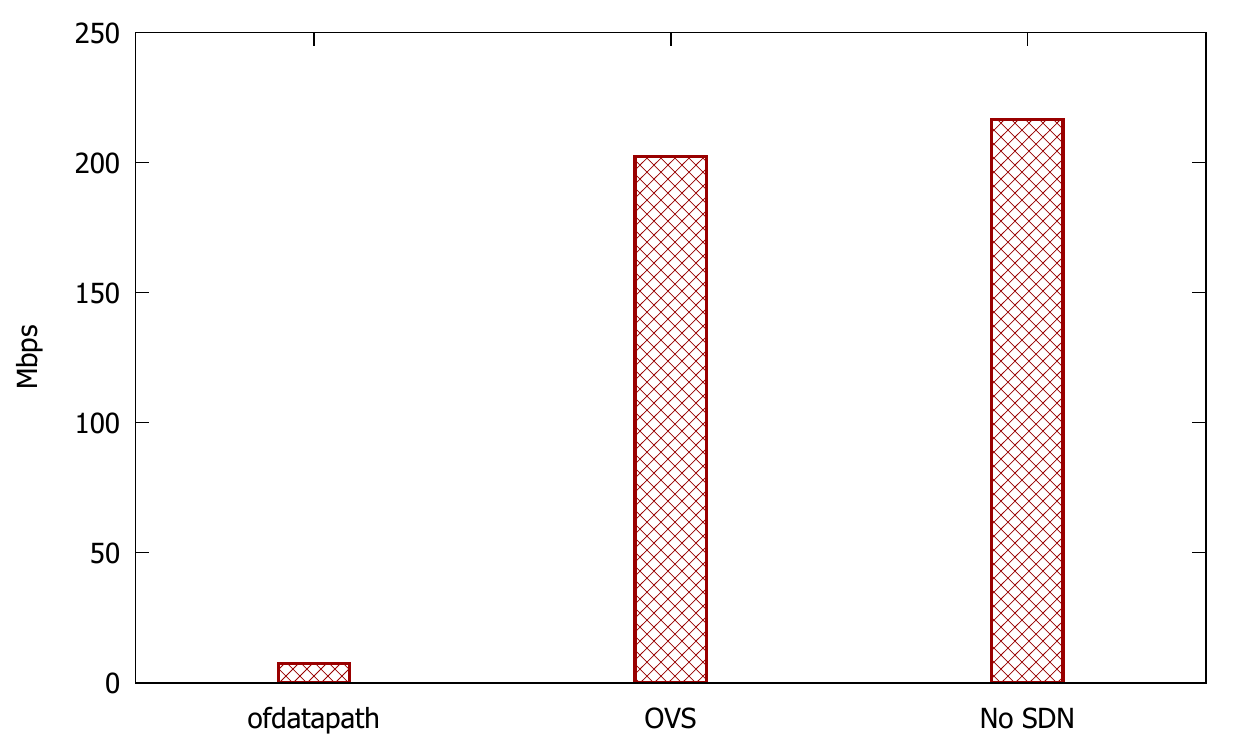}
  \caption{Switch Performance Results}
  \label{fig:Baseline-Results}
\end{figure}

Figure~\ref{fig:Baseline-Results} shows the maximum achievable throughput between the host and the server for three cases. It can be seen that OVS outperforms ofdatapath by a significant factor. This is not surprising since ofdatapath is a user-space implementation while OVS is a kernel implementation of an OpenFlow switch. We further observe that OVS incurs a minor performance penalty compared to the \emph{No SDN} case. This happens because we introduced an extra level of indirection with the virtual interface pair \emph{veth0} and \emph{veth1}, and the extra processing that this requires.
Due to the poor performance of ofdatapath, we decided to only consider OVS for our remaining experiments.

\section{Evaluation for Static Link Capacity}
\label{sec:Evaluation for Static Link Capacity}

In this section, SDN load balancing approaches will be evaluated when the capacity of the links is predefined and fixed. The evaluation is carried out in two ways: with a uniform and a non-uniform link capacity.

\subsection{Uniform Link Capacity}
\label{subsec:Uniform Link Capacity}
\begin{figure}[t]
  \centering
   \includegraphics[width=3.35in]{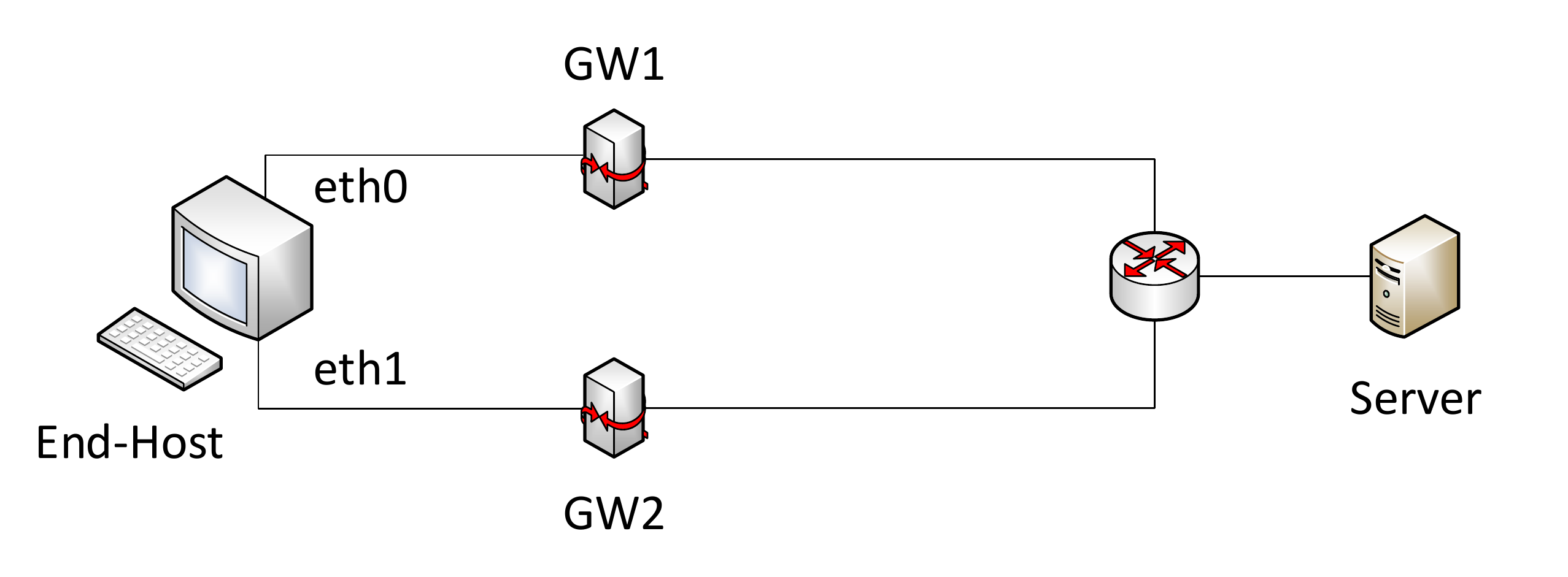}
  \caption{Load Balancing Experiment Scenario}
  \label{fig:Network-Topo}
\end{figure}

For the next experiment, we considered the scenario shown in Figure~\ref{fig:Network-Topo}, where a host is equipped with two network interfaces (\emph{eth0} and \emph{eth1}), each connected to a corresponding gateway (\emph{GW1} and \emph{GW2}), both of which are connected to another router, which is then connected to a web server. We configure both host interfaces with a uniform capacity of 10 Mbps \footnote{All other interfaces and links do not have a capacity limit.}.

At the host, we generate 100 HTTP GET request for a file located on the server, and we measure the time for the completion of these 100 downloads.

Each HTTP request results in the establishment of a new TCP connection, which we can distribute over the two available interfaces using our SDN-based load balancing methods.

\begin{figure}[t]
  \centering
  \includegraphics[width=3.35in] {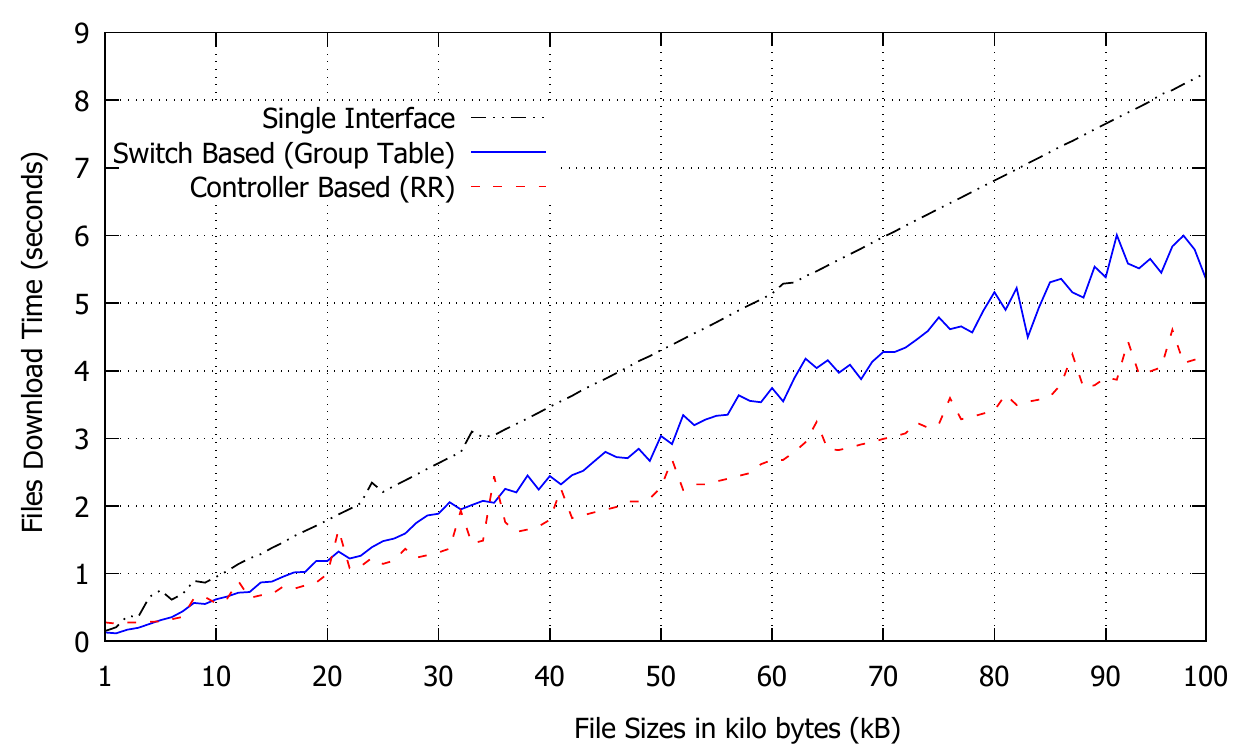}
  \caption{100 Files download time with respect to increasing file size (kB/file)}
  \label{fig:UniformLinkCapacity}
\end{figure}

Figure~\ref{fig:UniformLinkCapacity} shows the measured time for the 100 file downloads for various file sizes, ranging from 1kB to 100kB. 
The figure shows 3 graphs. The \emph{single interface} graph serves as a baseline and shows the download time if only a single host interface is used.

As expected, the time is (roughly) linear with the file size. For a file size of 100 kB, the download time when using a single interface is 8.4 s. 
The corresponding time for our controller-based load balancing (Round Robin) is 4.2 s, i.e half of the single interface case. 
This means our controller-based load balancing approach optimally utilises the aggregated link capacity of the two interfaces.

However, we notice that our switch-based load balancing approach (Group Table) does not perform as well, and achieves a download speed of significantly less than a factor of 2. 
This happens because we used the \emph{ab} traffic generation tool with a \emph{concurrency level} $C= 2$, which means that only two threads or processes are used to generate the requests, without pipelining. This works fine for a Round Robin mechanism, where the interface choice alternates between the two available options. In contrast, the switch-based mechanism relies on a hash function and the interface choice is pseudo-random; therefore, the same interface can be chosen multiple times in a row. With a concurrency level of 2, this results in one request in the queue while the second interface is idle. 

\begin{figure}[t]
  \centering
    \includegraphics[width=3.35in] {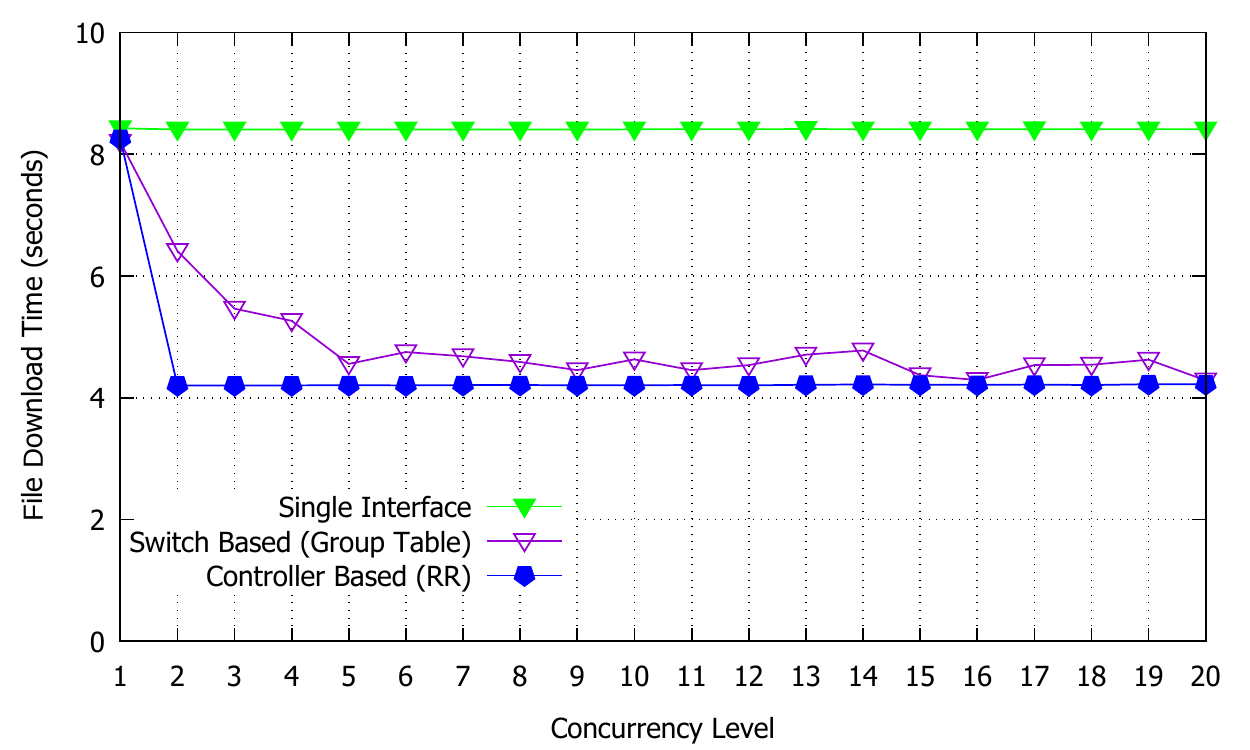}  
  \caption{Impact of concurrency levels upon the file download time}
  \label{fig:ConcurrencyLevels}
\end{figure}

We investigated various concurrency levels on the proposed load balancing approaches to figure out the suitable level that can satisfy maximum efficiency gain across the links/interfaces. Figure~\ref{fig:ConcurrencyLevels} shows the results of our experiment, where we measured the download time for a range of concurrency levels, for a fixed file size of 100kB. It is clear that for the Round Robin load balancer, the maximum efficiency gain is reached for $C=2$ already. The hash-based load balancer converges towards the optimal gain with an increasing value of $C$.  The Round Robin load balancer, therefore, has an advantage in scenarios with a small number of parallel requests.
To exclude this factor in our comparison, we choose a value of $C=20$ in our following experiments.

%

\begin{figure}[t]
  \centering
  \includegraphics[width=3.35in] {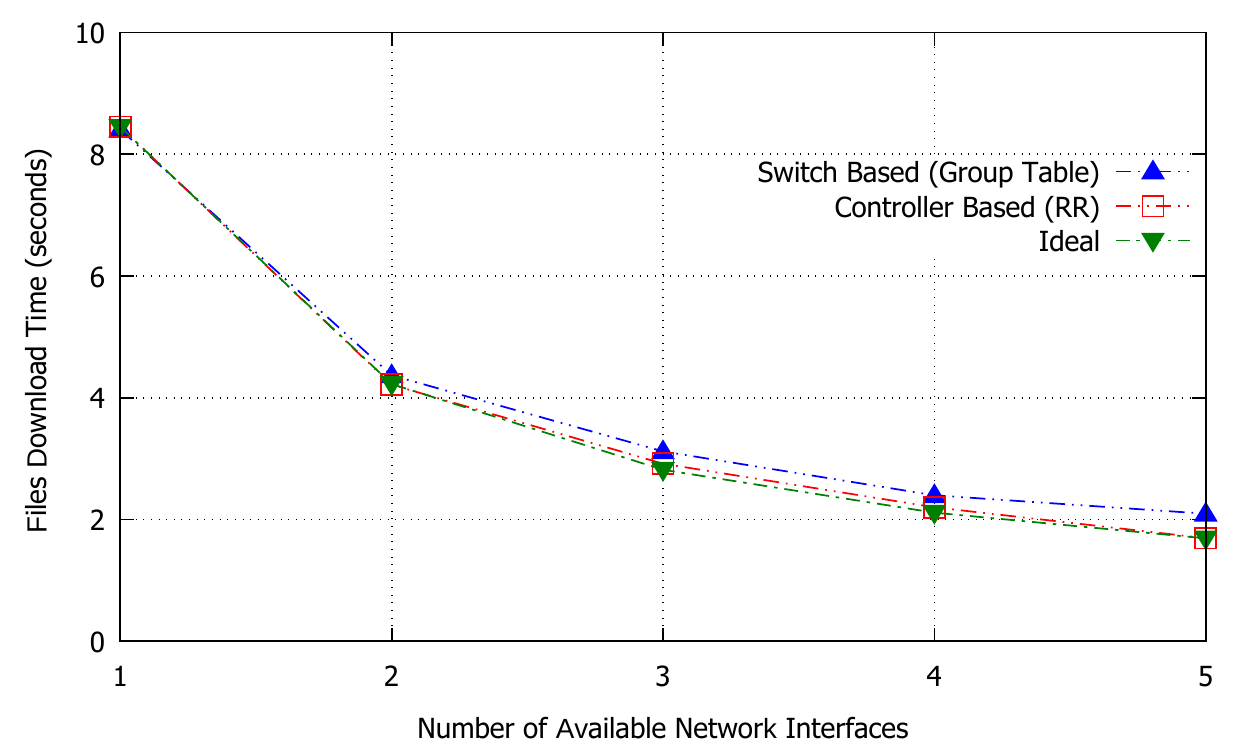}
  \caption{Effect of number of the available network interface connections over the file download time}
  \label{fig:MultipleInterfaces}
\end{figure}

We  performed further experiments with a higher number of host interfaces ($N$), i.e. with $N= 3, 4$ and $5$.
We used a fixed file size of 100kB for these experiments and 100 HTTP requests, as previously.
Figure~\ref{fig:MultipleInterfaces} summarises the results and shows the download time for both the controller-based and the switch-based load balancing approaches for $N=1, 2, 3, 4, 5$.
We also included a graph that represents the ideal load balancer as a reference, where the download time decreases with $1/N$.

We can observe that the controller-based load balancer achieves very close to optimal efficiency, with only a maximum difference of 0.5\% from the ideal case.
The switch-based approach also performs very well, but with a slightly increased gap to the optimal performance, with a maximum difference to the ideal load balancer of around 6\%.

\subsection{Non-Uniform Link Capacity}
\label{sub:Non-Uniform Link Capacity}

We now consider the same scenario as shown in Figure~\ref{fig:Network-Topo}, but this time with non-uniform link capacities, i.e. \emph{eth0} has 10 Mbps and \emph{eth1} has 20 Mbps capacity. 
Both our controller-based (Round Robin) and switch-based (Group Tables with hash-based selection) 
 load balancing mechanism performs equal load sharing, where each interface is assigned the same amount of traffic. This is not ideal in a case with non-uniform link capacities. 

Hence we use the weighted load balancing approaches with the scenario shown in Figure~\ref{fig:Network-Topo}. While the WRR algorithm is used instead of the RR for the controller based approach, the Weighted Group Table is used instead of a simple Group Table as a switch-based traffic distribution approach.

\begin{figure}[t]
  \centering
   \includegraphics[width=3.35in] {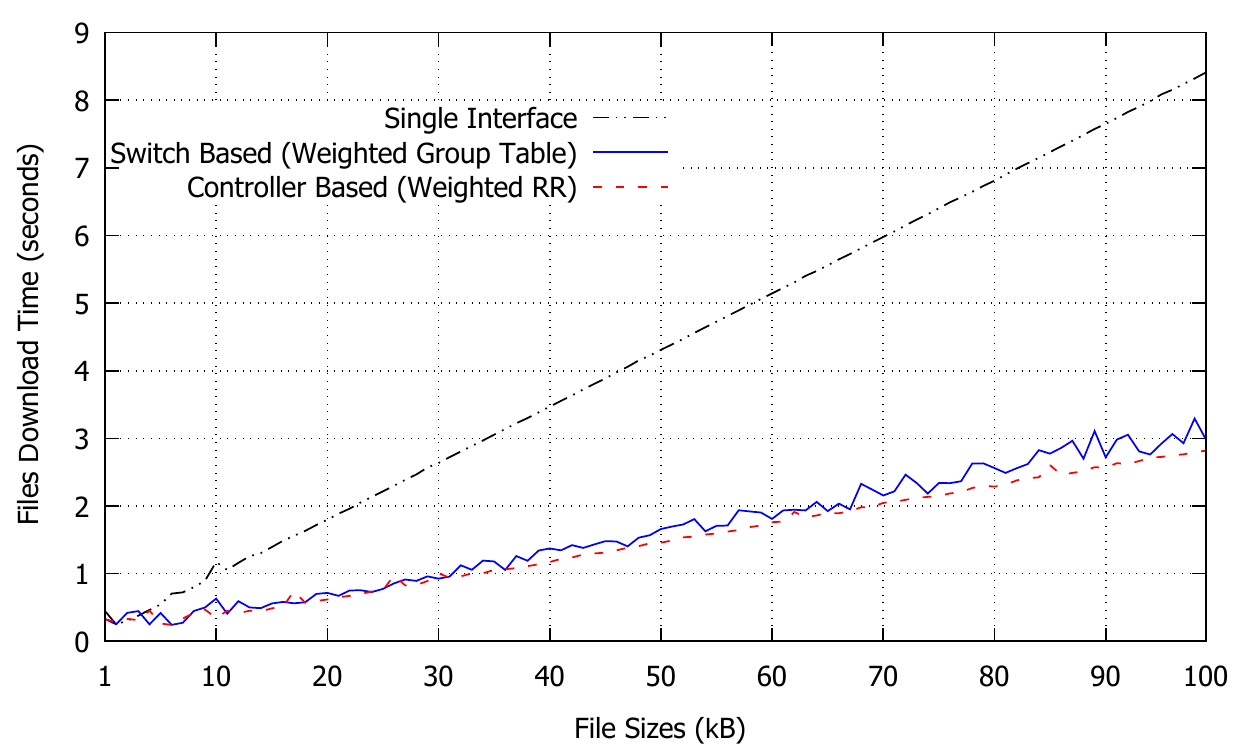}
  \caption{Files download time with respect to different file sizes (kB/file) in case of non-uniform link capacity}
  \label{fig:NonUniformLinkCapacity}
\end{figure}

Figure~\ref{fig:NonUniformLinkCapacity} shows results of our download time measurement for a 100 HTTP requests. 
As before, we use file sizes ranging from 1kB to 100kB, and we observe that the file download time increases linearly with growing file size.
As a reference, we also included the download time for the case of a single interface with 10 Mbps capacity.
The aggregated capacity of both interfaces is 30 Mbps, and we would expect an ideal load balancer to achieve a download time reduction by a factor of 3.

As mentioned previously, the download time for 100 kB files via the single 10 Mbps interface is 8.4 s. The ideal download time for an aggregated capacity of both interfaces is a third of that, i.e. 2.8 s.
We can see that the controller-based load balancer with a download time of 2.82 s comes very close to the optimal.
The switch-based load balancer, using a weighted hash-based approach, achieves a slightly higher time of 2.99 s, but still less than 7\% of the optimal value.

We also observe a slightly higher variability of switch-based approach compared to the controller-based. 
We attribute this to the pseudorandom interface selection (via a hash function) of the switch-based approach, compared to the deterministic selection of the Weighted Round Robin approach in the controller-based approach.


In summary, the SDN-based load balancing approaches perform very well when the capacity of the links is static. In the next part of our experiments, the load balancing performance will be examined when the capacity is periodically changing.

\section{Evaluation for Dynamic Link Capacity}
\label{sec:Evaluation for Dynamic Link Capacity}

In a real wireless network, the capacity of last hop communication link changes frequently due to several factors. These include the distance from the access node, channel interference as well as the number of users sharing the channel. Thus the variation in the capacity should be taken into account to make a better decision about traffic distribution.

It is therefore essential to estimate the capacity in advance to optimally distribute the traffic across the available links. An important approach to estimate the capacity is to use Variable Packet Size (VPS) probing adaptive to the concept of SDN~\cite{al2016link}.

Briefly, the probing is performed on the last hop connection\footnote{In our topology shown in Figure \ref{fig:Network-Topo}, the last hop connections are the directly connected links to the End-host, which are \textit{eth0-GW1} and \textit{eth1-GW2} links.} presuming that the bottleneck lies there. The SDN controller crafts different size probing packets and sends them to the OpenFlow switch. The latter forwards the packets via a specific network interface to probe a directly connected gateway. When the replied packets reach back to the switch, they will be forwarded to the controller. The controller picks the least Round Trip Time (RTT) packet of different size groups presuming that at least there is one packet of each group which has either experienced some marginal delay or not experience any delay at all. Then, the capacity is estimated by the controller using the linear regression technique on the least delayed packet.

Once the capacity of the links has been figured out, it will be leveraged by the SDN controller to decide how much traffic should be allocated to each link. Although this is considered a realistic approach for achieving dynamic load balancing, it is difficult to provide a repetitive controlled environment for evaluating the results. Thus, we first measure the bandwidth of the available links and store the results as a data-set to be used later in our system.

In our implementation, the measured data-set is stored on the directly connecting gateways (GW1,GW2) as well as on the end-host of the proposed topology (shown in Figure~\ref{fig:Network-Topo}). At gateways, the bandwidth data-set is used to emulate the link bandwidth variation over time whereas it is utilised by the SDN controller to proportionally allocate network flows over the links. This is to make sure that the bandwidth values should be synchronised between the connecting nodes. 

To achieve such synchronisation, a client-server socket connection is established between the gateways and the end-host. The controller triggers the binding process by sending messages to the gateways. Then a continuous process runs simultaneously with the load balancing mechanism. The process encompasses reading and assigning the bandwidth measurements on the last hop links\footnote{Since the traffic direction based on the experimental traffic generation tool, ab, is from the server to the client, binding the bandwidth data-set should be done on the gateway interfaces.}, as well as reading and leveraging the same measurements by the SDN controller.

During the evaluation of the proposed load balancing algorithms, the throughput of the transferred traffic should be measured. This is important to compare the performance of the proposed load balancing algorithms with the emulated bandwidth data-set. To frequently measure throughput, the OpenFlow \textit{Port Stats} message is sent by the SDN controller to probe statistical information of OpenFlow switch ports, namely: \textit{Eth0}, \textit{Eth1} and \textit{Veth0}. While the frequent probing of the first two interfaces represents throughput of the emulated links, probing the virtual interface represents the aggregated throughput of both links.

We also evaluate our load balancing algorithms with MPTCP protocol that we described earlier in Section \ref{sec:Background}. The protocol has been configured on the end-host and the server of the proposed topology shown in Figure~\ref{fig:Network-Topo}. For the configuration of MPTCP, the default settings were used as stated in MPTCP official website \cite{paasch2013multipath}. Likewise, with our load balancing algorithms, we also measured the throughput on each link. However, we cannot use \textit{Port Stats} messages to probe the link statistics since the SDN paradigm is not applicable in such a scenario. Instead, we used \textit{ifconfig}, a Linux command line tool for network interface configuration \cite{ifconfig}. The command is leveraged as a part of a background process that collects various statistics about interfaces per second basis, including the number of sent/received bytes. Moreover, the aggregated throughput is calculated via accumulating the throughput of the last hop links which differs from computing the aggregated throughput with our proposed system.

After explaining the testbed setup, two link bandwidth scenarios will be used to evaluate the load balancing algorithms. The first scenario uses synthetic bandwidth data-set, while the second uses realistic bandwidth measurement.

\subsection{Simple Dynamic Capacity}
\label{subsec:Simple Dynamic capacity(Step Function)}

This experiment aims to move our evaluation gradually from static to the dynamic link capacity case. We presumed that only one link has a simple variation in bandwidth. In this scenario, we created a synthetic data-set of bandwidth that will be leveraged later for evaluating different load balancing algorithms using the architecture in Figure~\ref{fig:Network-Topo}.

\begin{figure}[t]
  \centering
  \includegraphics[width=3.35in]{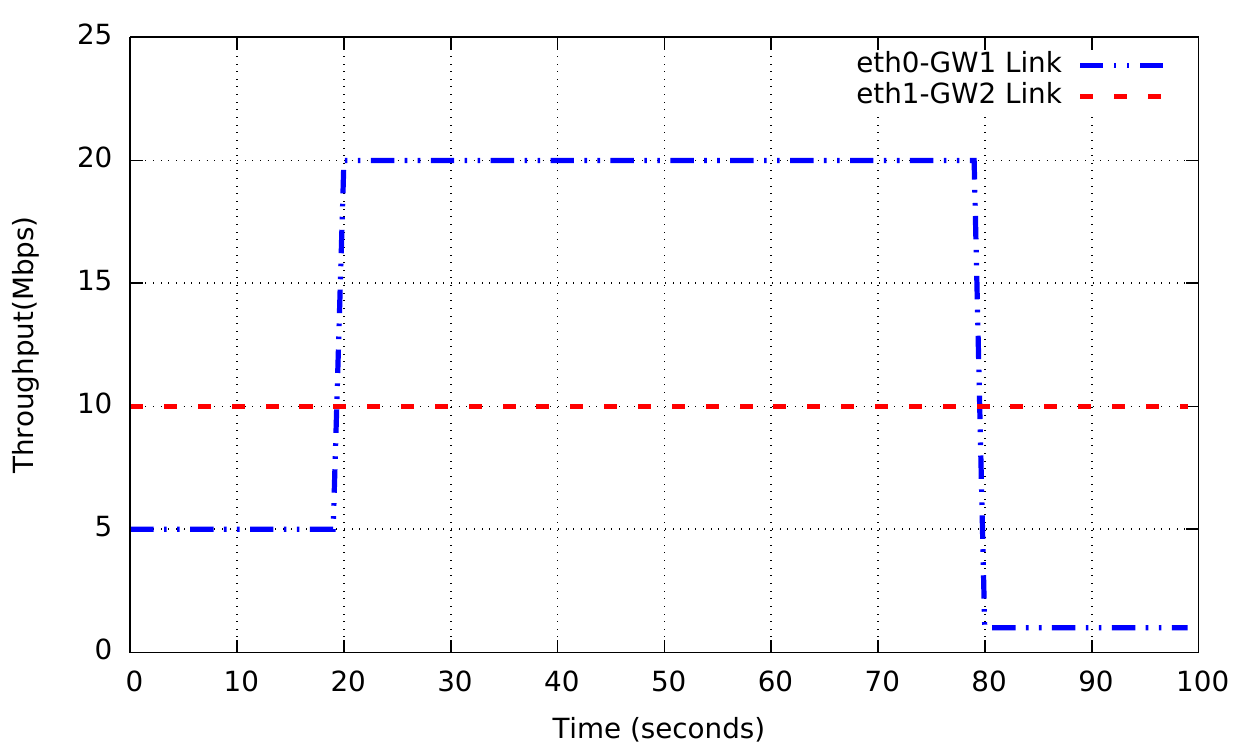}
  \caption{Available synthetic bandwidth of the two network links over time}
  \label{fig:Emulated BW}
\end{figure}

Figure~\ref{fig:Emulated BW} shows the bandwidth data-set of the directly connected links to the End-host over 100 seconds. The assigned bandwidth to emulated \textit{eth1-GW2} link is fixed to 10 Mbps, while the bandwidth of \textit{eth0-GW1} link varies among three levels. The link bandwidth starts at 5 Mbps and remains there for 20 seconds. Then, it levels off to 20 Mbps for the 60s before dropping to 1 Mbps and continues till the end of the experiment.

\par
\begin{figure}[t]
  \centering
  \captionsetup{justification=centering}
  \includegraphics[width=3.35in]{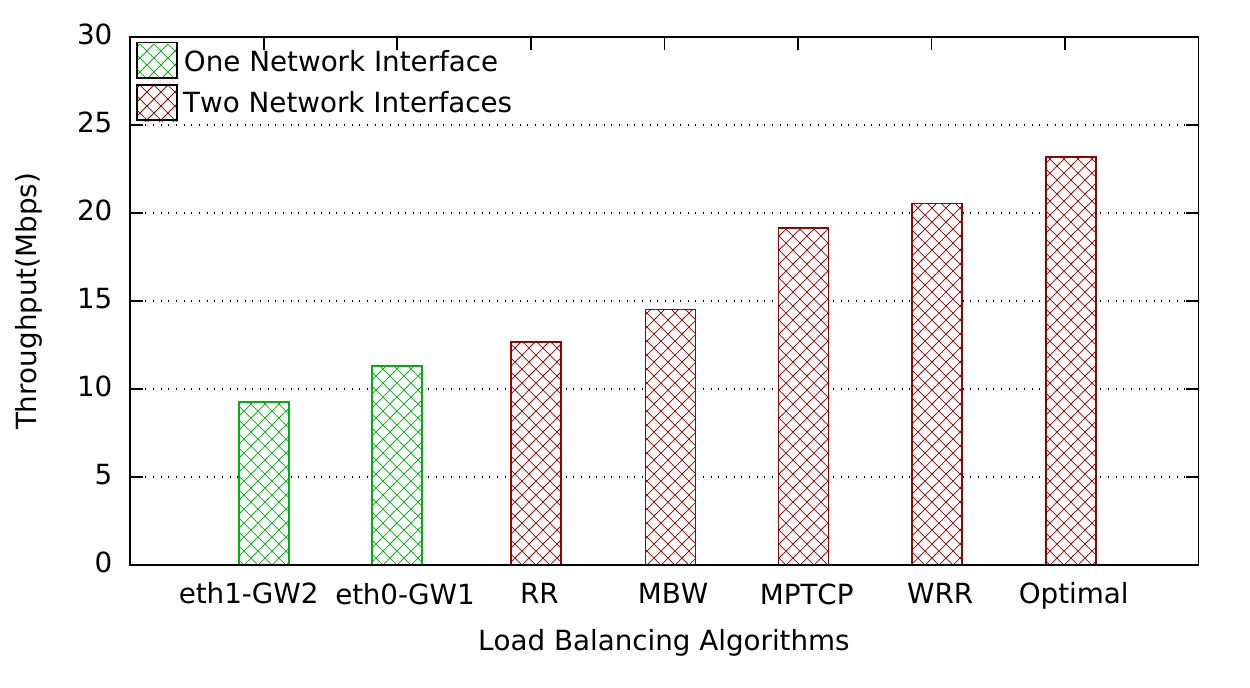}
  \caption{Throughput of different load balancing algorithms\\(Simple dynamic capacity scenario)}
  \label{fig:OVS-Algorithm}
\end{figure}

For the experimental setup, $tc$ tool was used to bind the bandwidth per sec basis on the directly connected links to the end-host. Furthermore, the network traffic was generated using $ab$ via sending HTTP requests with concurrency level 20 to the 100KB file size on the server. During the evaluation, it has been noticed that the group tables, related to the Switch-Based load balancing approach, could not cope with such emulated bandwidth. The approach has stopped in the first quarter of the experiment time. Hence, the evaluation has been conducted using controller-based approaches only.

Figure~\ref{fig:OVS-Algorithm} shows the achieved throughput by applying various load balancing algorithms on SDN based end-host. The experiment shows the throughput in two cases: by utilising one network interface (with simple traffic forwarding) and by utilising two network interfaces (with load balancing). We also evaluated our load balancing approaches against the performance of MPTCP protocol as a benchmark and compared them with the \textit{optimal} throughput that represents the addition of the synthetic bandwidth on both links.

The throughput of transferring the network traffic via one link is around 9Mbps when the eth1-GW2 link is used and slightly above 11Mbps when eth0-GW1 link is used. However, when both links are used, the achieved throughput is higher.

As shown in Figure~\ref{fig:OVS-Algorithm} RR algorithm achieves throughput (approximately 12.5Mbps) which is more than using one link, but the performance is still far less than optimal ($\approx$ \%45). This is because RR equally allocates the network flows to different capacity links; therefore, optimal link bandwidth utilisation does not occur.

\begin{figure}[t]
  \centering
  \captionsetup{justification=centering}
  \includegraphics[width=3.35in]{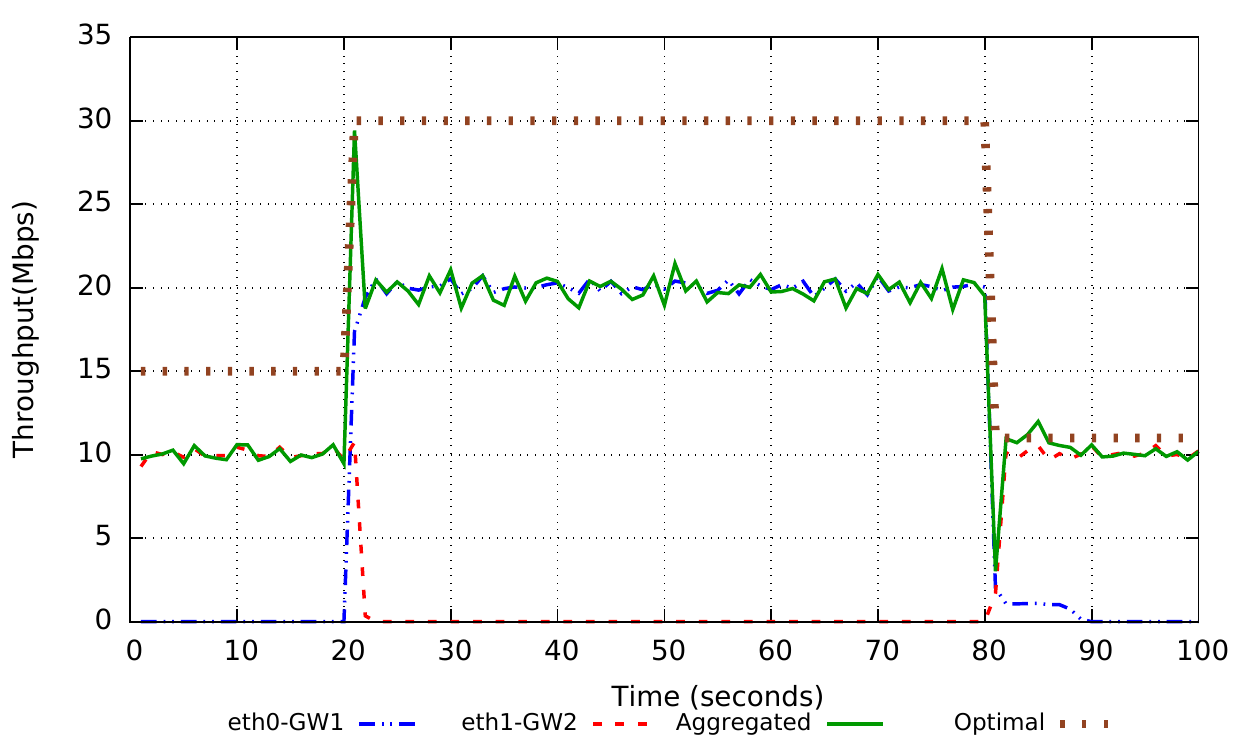}
  \caption{Throughput of the last hop links using MBW\\(Simple dynamic capacity scenario)}
  \label{fig:MBW Traffic Flow}
\end{figure}
\par
\begin{figure}[t]
  \centering
  \captionsetup{justification=centering}
  \includegraphics[width=3.35in]{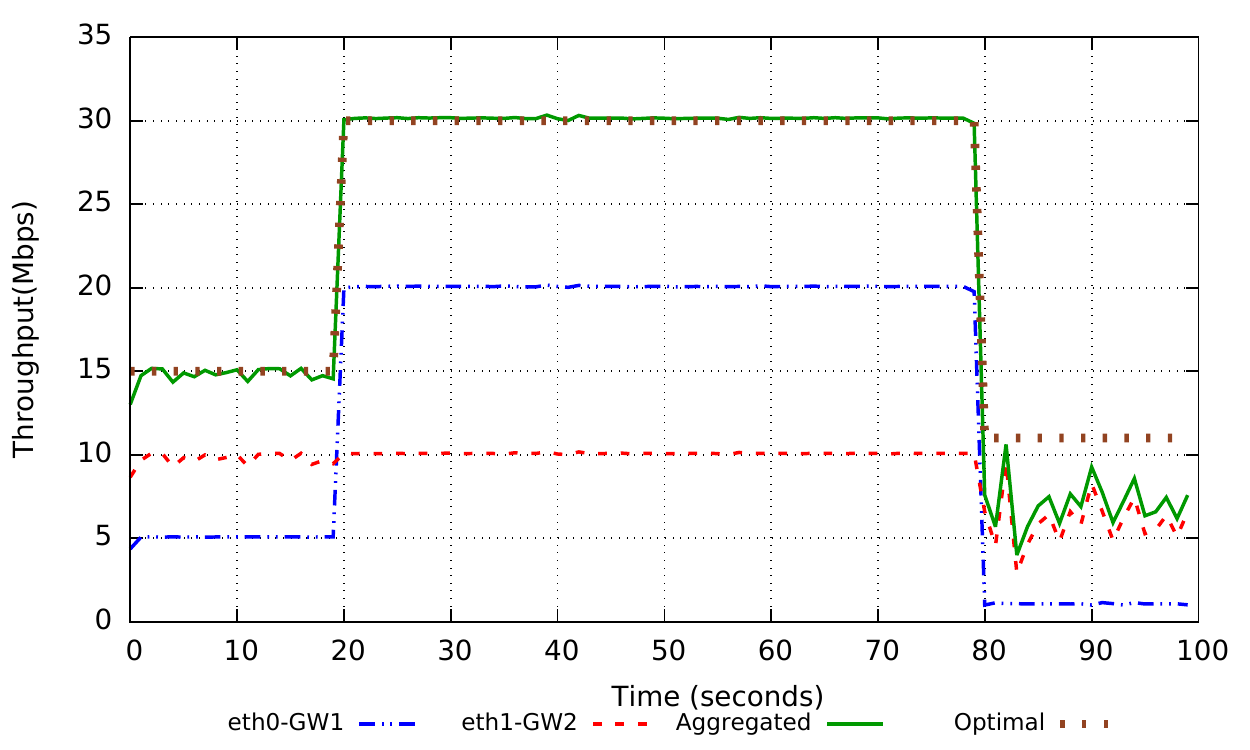}
  \caption{Throughput of the last hop links using MPTCP\\(Simple dynamic capacity scenario)}
  \label{fig:MPTCP Traffic Flow}
\end{figure}

In the case of MBW algorithm, the achieved throughput is about 14.5Mbps which is higher than the throughput of RR. We investigated the performance of this algorithm via measuring the throughput of OVS ports using \textit{Port Stats} OpenFlow messages as described earlier in this section. Figure~\ref{fig:MBW Traffic Flow} shows the throughput of the last hop links gathered over 100s and the Aggregated throughput of both links.  It is seen that the Aggregated throughput is almost the same as the maximum bandwidth link besides two sudden spikes that happened when eth0-GW1 link levels off at 20Mbps and drops to 1Mbps. The first spike happens due to the temporary utilisation of both network interfaces when the algorithm selects another link with maximum bandwidth as previously explained in Section \ref{sec:loadBalancingApproaches} of this paper as well as in a previous study \cite{rahmati2010seamless}. The second spike happens due to the sudden drop on the eth0-GW1 link leading Apache tool to resolve the status of the sent packets (via either waiting to be acknowledged or to drop the connection) before generating new flows that will be allocated on the eth1-GW2 link. Overall, MBW mostly utilises the link with maximum bandwidth optimally, which is why it outperforms RR algorithm that utilises multiple network links, simultaneously.


Regarding the use of MPTCP, it achieves throughput about 19Mbps which is better than the previous algorithms. However, it is around 17.5\% less than optimal. To investigate the performance of this algorithm, we measured the throughput of MPTCP links using \textit{ifconfig} Linux command as previously explained. Figure~\ref{fig:MPTCP Traffic Flow} shows throughput over 100 seconds of the directly connected links to the end-host when MPTCP is used. We noticed that throughput of both links almost matches the optimal case except there is a fluctuation below the optimal throughput happened around 80th second and onwards. This is due to the drop happened in the bandwidth of the eth0-GW1 link that even affects the performance of the eth1-GW2 link. Note that the MPTCP implementation uses a decoupled congestion control algorithm (e.g. cubic~\cite{paasch2013multipath}) that maintains separate congestion windows for each of the two sub-flows. As a result, when congestion occurs on one link, the algorithm fails to push more traffic over the uncongested path~\cite{ford2016tcp}, which subsequently affects the resulting throughput on the eth1-GW2 link.
 
\begin{figure}[t]
  \centering
  \captionsetup{justification=centering}
  \includegraphics[width=3.35in]{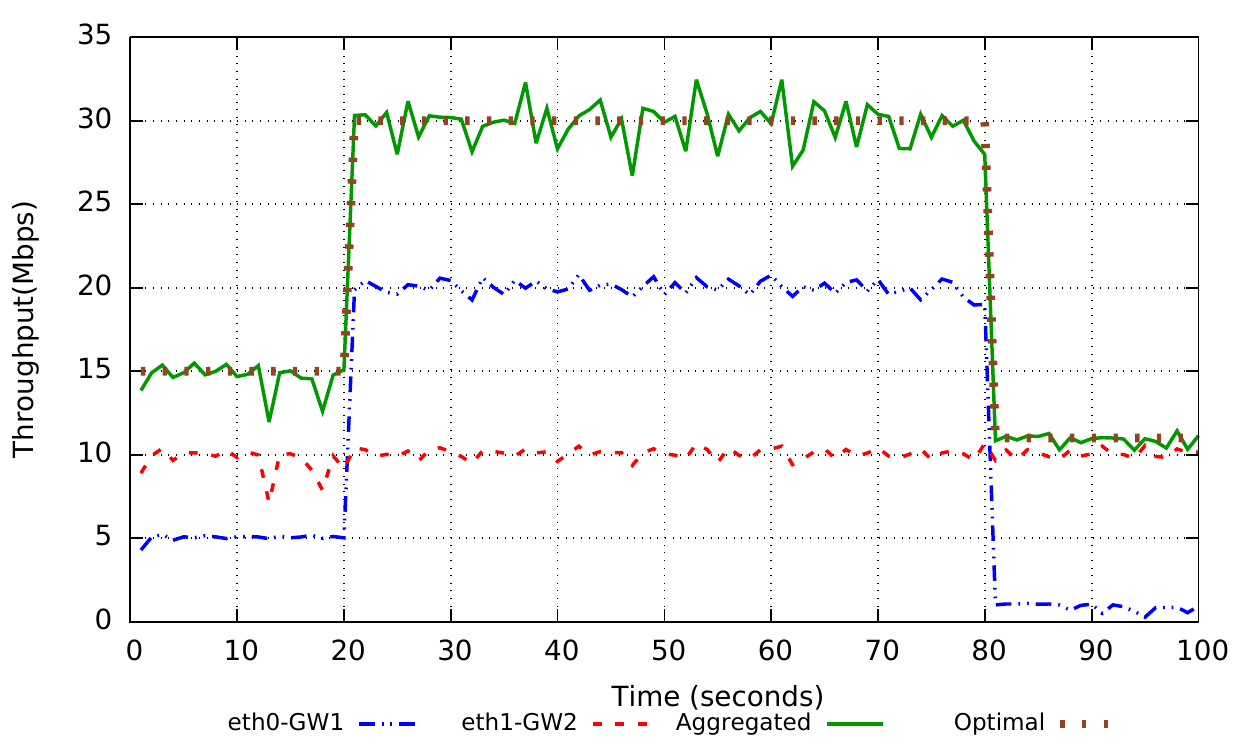}
  \caption{Throughput of the last hop links using WRR\\(Simple dynamic capacity scenario)}
  \label{fig:WRR Traffic Flow}
\end{figure}

In the end, WRR achieves more than 20.5Mbps throughput, outperforming the previously mentioned algorithms in particular MPTCP. This is because WRR allocates TCP flows based on the computed links' weights. This can be verified from the throughput per link as shown in Figure~\ref{fig:WRR Traffic Flow}. Also from the figure, it can be clearly seen that bandwidth of the links is optimally utilised without any drop.

\begin{figure}[t]
  \centering
  \includegraphics[width=3.35in]{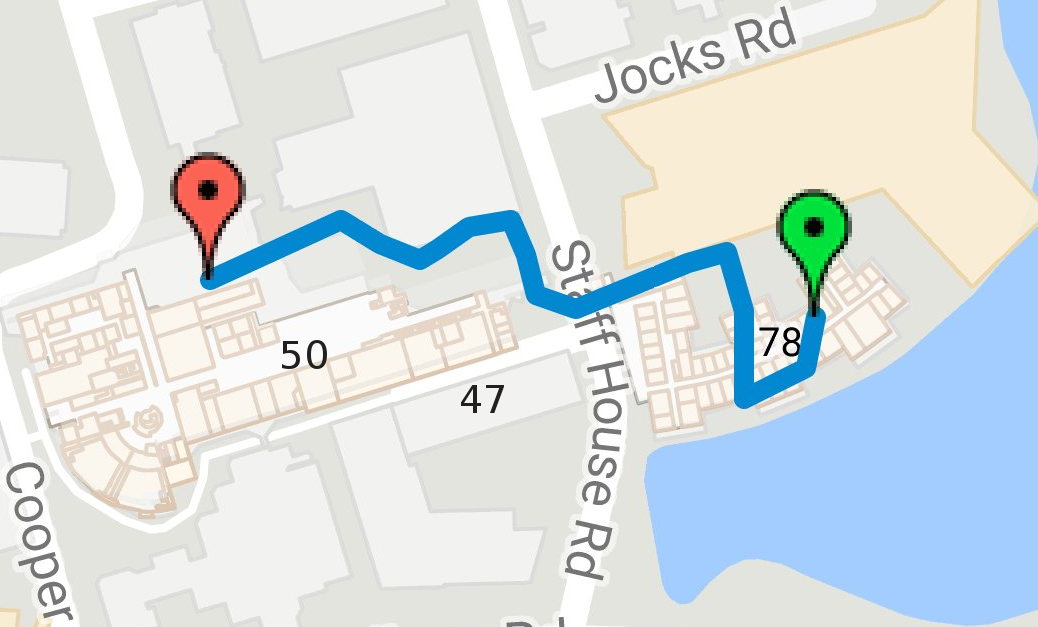}
  \caption{Bandwidth measurement path at UQ}
  \label{fig:UQ-BW-Engineering-Map}
\end{figure}

\begin{figure}[t]
  \centering
  \includegraphics[width=3.35in]{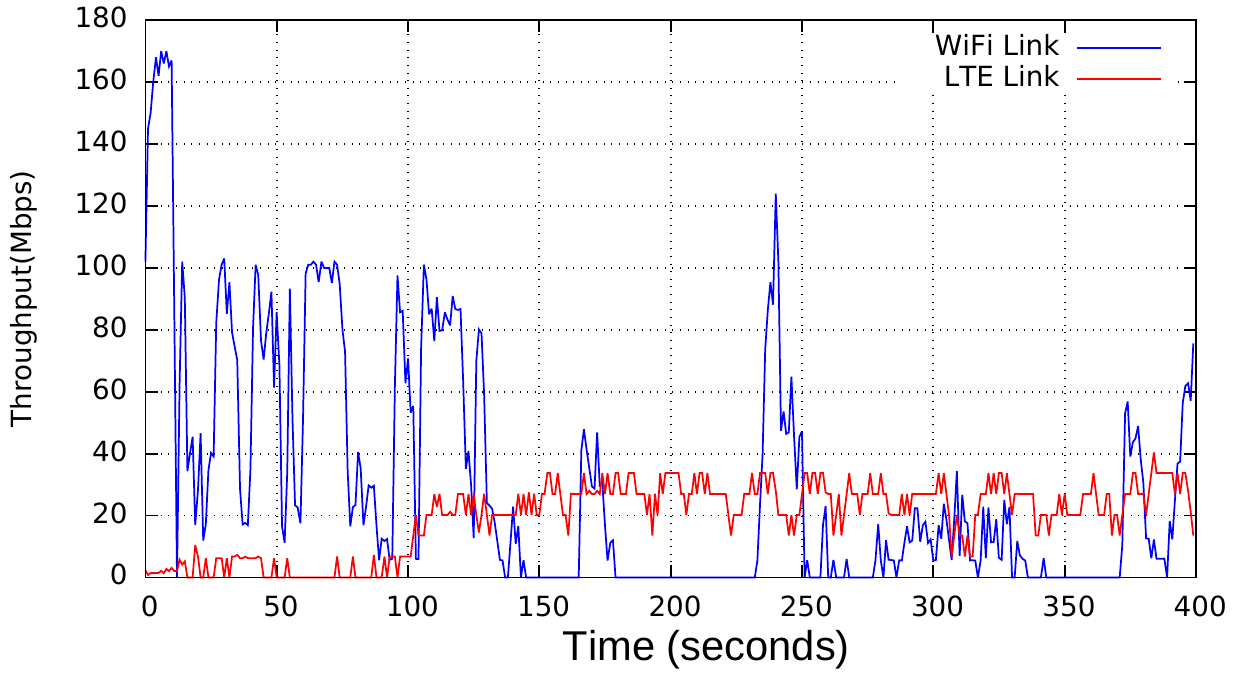}
  \caption{UQ bandwidth measurement}
  \label{fig:UQ-EmulatedBW}
\end{figure}

To summaries, our controller based load balancing approaches, in particular WRR, do not only achieved better performance compared with using a single network interface, but also outperform the MPTCP. In the following scenario, we will evaluate the performance of our controller based end-host approaches with a real bandwidth measurement of WiFi and 4G networks.

\subsection{Realistic scenario}
\label{subsec:Realistic scenario}

In this scenario, a realistic bandwidth measurement is used to evaluate the proposed load balancing approaches. 

The measurement was conducted in The University of Queensland (UQ) St. Lucia campus. The experimenter was following a certain path shown in Figure~\ref{fig:UQ-BW-Engineering-Map} at the university to collect the bandwidth measurement. In the experiment two laptops were used to measure the bandwidth of WiFi and LTE links independently via Ipref tool \cite{iperf3}. The experiment started indoor of building 78 and the experimenter went outdoor to finish the experiment at building 50 as seen from Figure~\ref{fig:UQ-BW-Engineering-Map}. Further relevant details of our experiment are explained in our previous work~\cite{al2017flow}.

Figure~\ref{fig:UQ-EmulatedBW} shows the bandwidth measurement for WiFi and LTE links which endures for 400 seconds. Initially, when the experiment conducted indoor (which is up to the 100th second), the WiFi network shows good bandwidth performance after which it experiences several drops in the outdoor environment. In contrast, LTE has a low bandwidth in the indoor measurement and then recovered to remain around 20Mbps until the end of the experiment.

Before evaluating the proposed load balancing approaches, we would like to validate the dynamic change of the measured bandwidth in the testbed explained in Figure~\ref{fig:Baseline-Topology}. This is important to ensure that the dynamic behaviour of such bandwidth is reproducible in the proposed testbed. To emulate the measured bandwidth, \textbf{tc} tool is used to assign the bandwidth values on the directly connected gateway. Two traffic generation tools were used in the experiments. We used \textbf{Iperf} with time interval 1s in the baseline system, where SDN is not applied. Moreover, the \textbf{ab} is leveraged with SDN implementation\footnote{The reason for using ab with SDN is its ability to generate a multi-flow connection that cannot be done with Iperf tool}. To determine the throughput per second, the traffic which is passed through the \textit{Veth0} network interface was read each second using \textit{Port Stats} OpenFlow message.

\begin{figure}[t]
  \centering
  \includegraphics[width=3.35in]{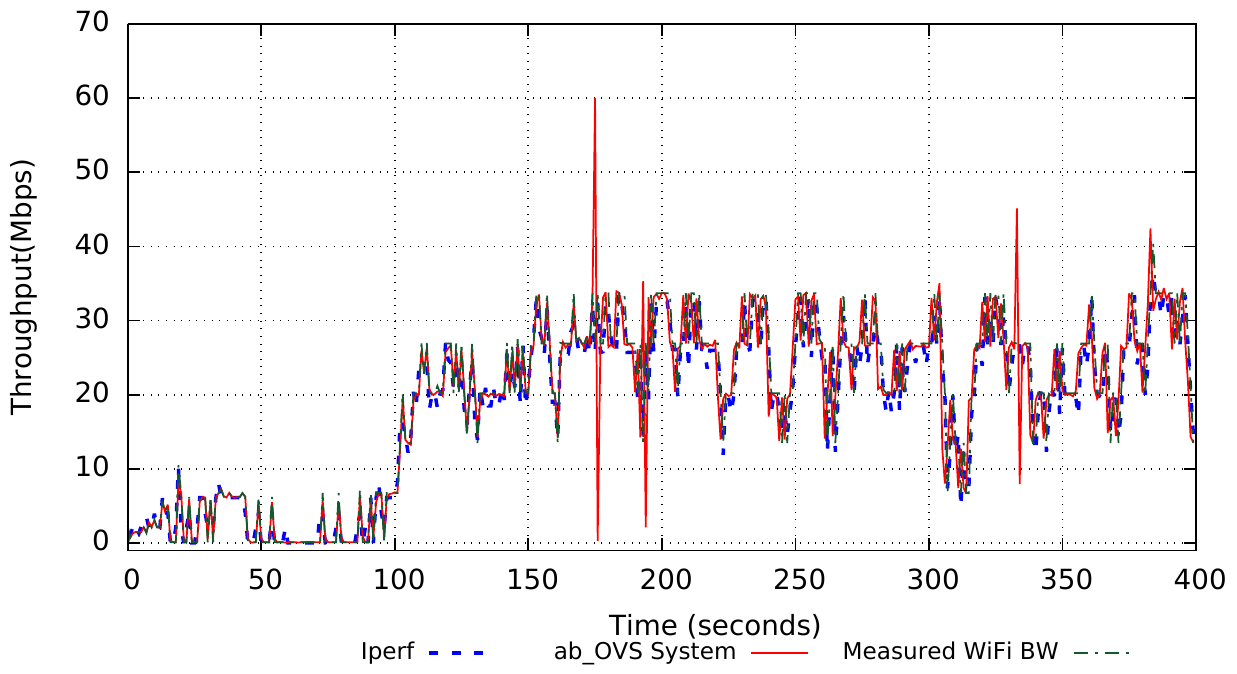}
  \caption{LTE measurement validation}
  \label{fig:UQ-LTE-Validation}
\end{figure}

\begin{figure}[t]
  \centering
  \includegraphics[width=3.35in]{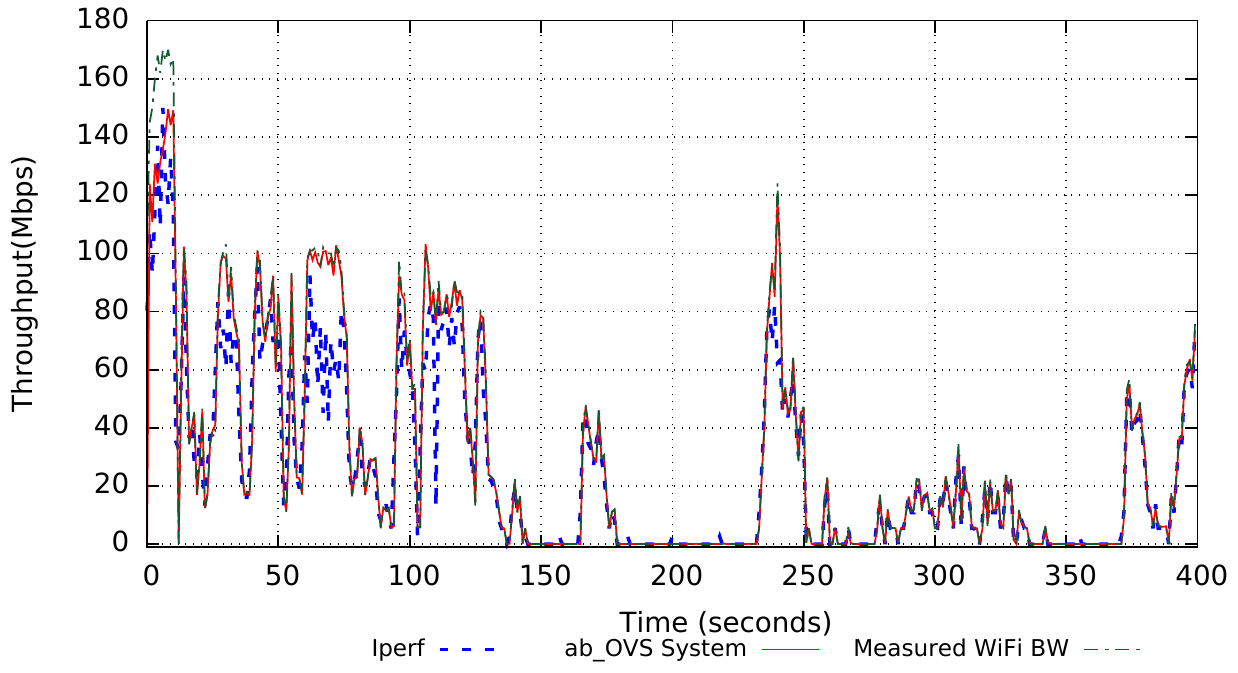}
  \caption{WiFi measurement validation}
  \label{fig:UQ-WiFi-Validation}
\end{figure}

Figure~\ref{fig:UQ-LTE-Validation} and \ref{fig:UQ-WiFi-Validation} show the validation of LTE and WiFi links respectively by measuring their throughput. The \textit{ab} and \textit{Iperf} tools, reflect almost the same trend of the measured bandwidth although the \textit{ab} tool have a sudden spike in LTE validation. In other words, the validation of the UQ bandwidth measurement seems acceptable especially using \textit{ab} tool with SDN implementation. The next step is to evaluate our traffic distribution algorithms with the validated measurement.

\begin{figure}[t]
  \centering
  \captionsetup{justification=centering}
  \includegraphics[width=3.35in]{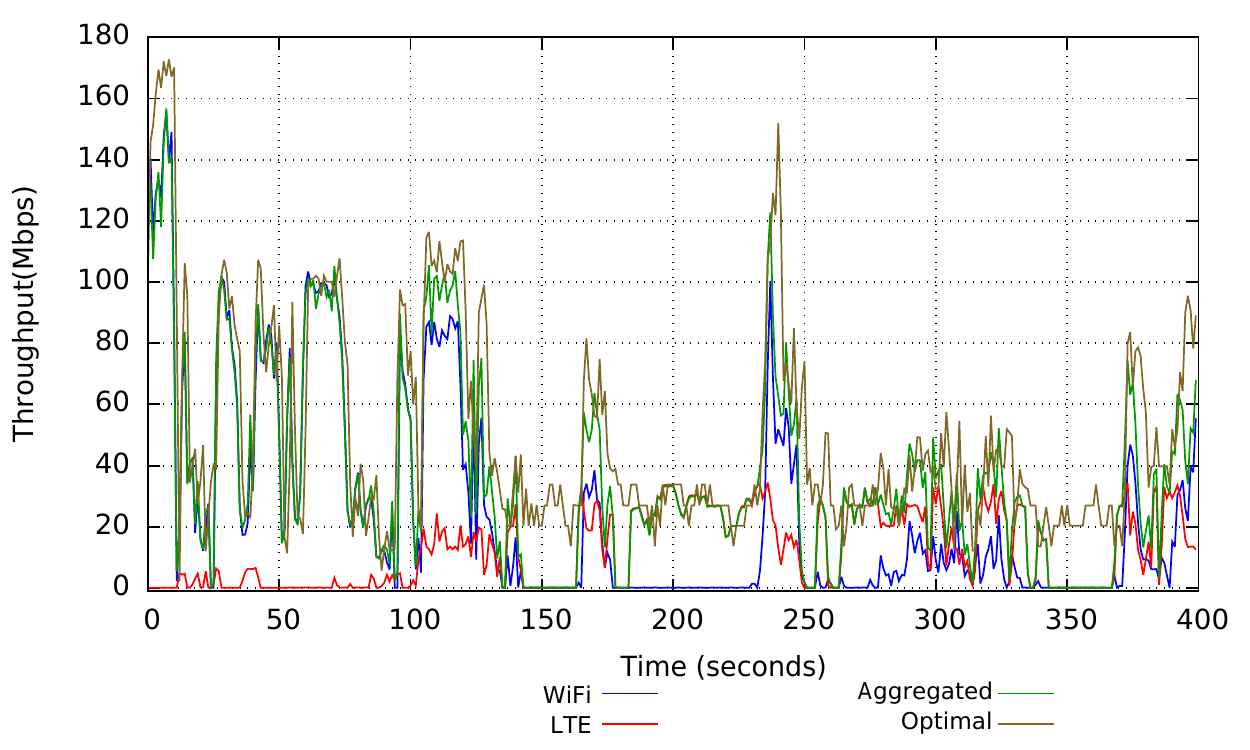}
  \caption{Throughput of the last hop links using WRR\\(Realistic dynamic capacity scenario)}
  \label{fig:UQ-WRRTrafficFlow}
\end{figure}

Figure~\ref{fig:UQ-WRRTrafficFlow} shows the throughput over 400 seconds of the proposed system, explained in Figure~\ref{fig:Network-Topo}, when WRR algorithm is used. It can be seen that the aggregated throughput is close to the optimal which is obtained via accumulating the measured bandwidth of WiFi and LTE links. However, there is a noticeable difference in the throughput between 150 to 350 seconds intervals. This could be due to the application socket timeout that maintains the connectivity. For instance, the Apache tool sends a request and the bandwidth of the carrying link is dropped suddenly. Consequently, the application is waiting to complete the request before sending another one. In this case, there is no traffic forwarding at that time until the bandwidth recovers again or the connection is reset. To investigate what is happening, the traffic forwarding using WRR is re-conducted with various socket timeout.

\begin{figure}[t]
  \centering
  \includegraphics[width=3.35in]{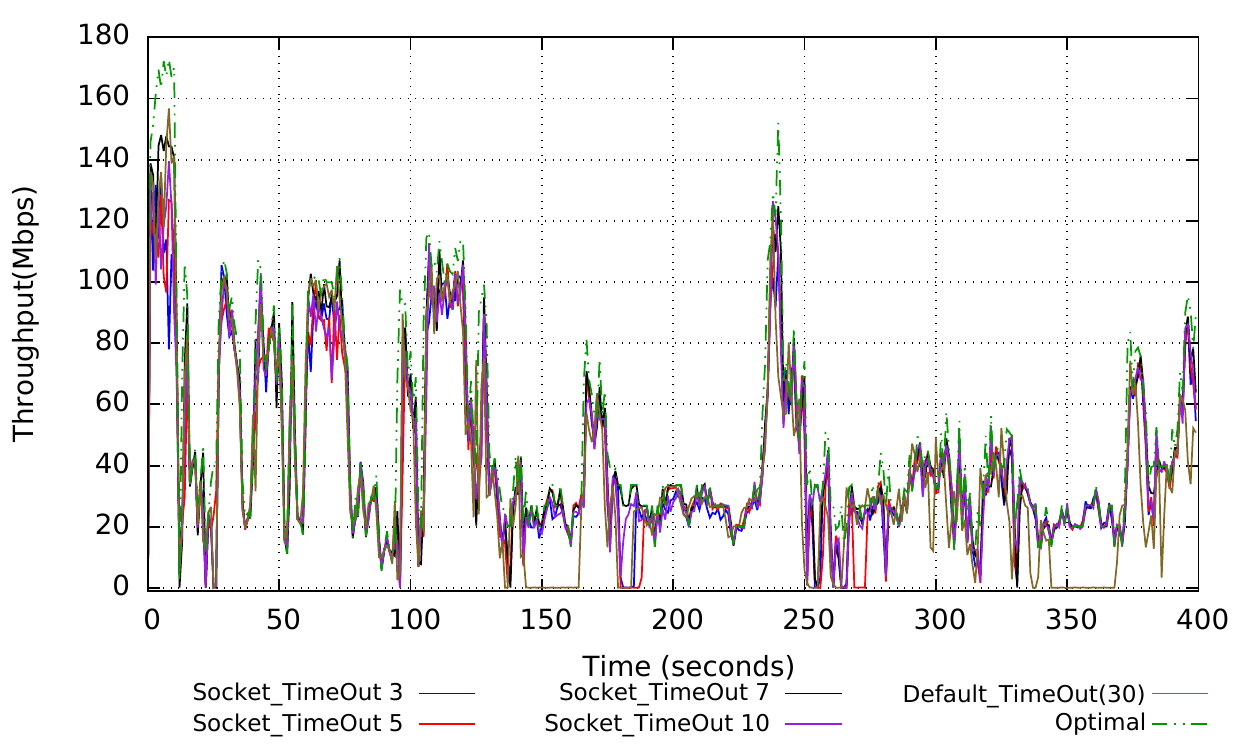}
  \caption{WRR aggregated throughput with different ab socket timeout}
  \label{fig:UQ-Apache_Socket_TimeOut}
\end{figure}

Figure~\ref{fig:UQ-Apache_Socket_TimeOut} depicts the aggregated throughput of WiFi and LTE over 400 seconds duration using WRR when different ab socket timeouts (3, 5, 7, 10 seconds and the default timeout of 30 seconds) are applied. With a small socket timeout, it is clear that the throughput does not drop as much as when the default timeout is applied. This is because, when a link bandwidth suddenly drops or becomes zero, the current request will be terminated based on the socket timeout, and the ab tool will create another TCP request that is likely to be allocated at the other link that has higher bandwidth. This explains that a small socket timeout achieves better throughput. To avoid the recurrent TCP flow re-establishment, that requires additional resources, and based upon our results, the socket timeout of 10 seconds is chosen for the rest of our experimental evaluations.

\begin{figure}[t]
  \centering
  \captionsetup{justification=centering}
  \includegraphics[width=3.65in]{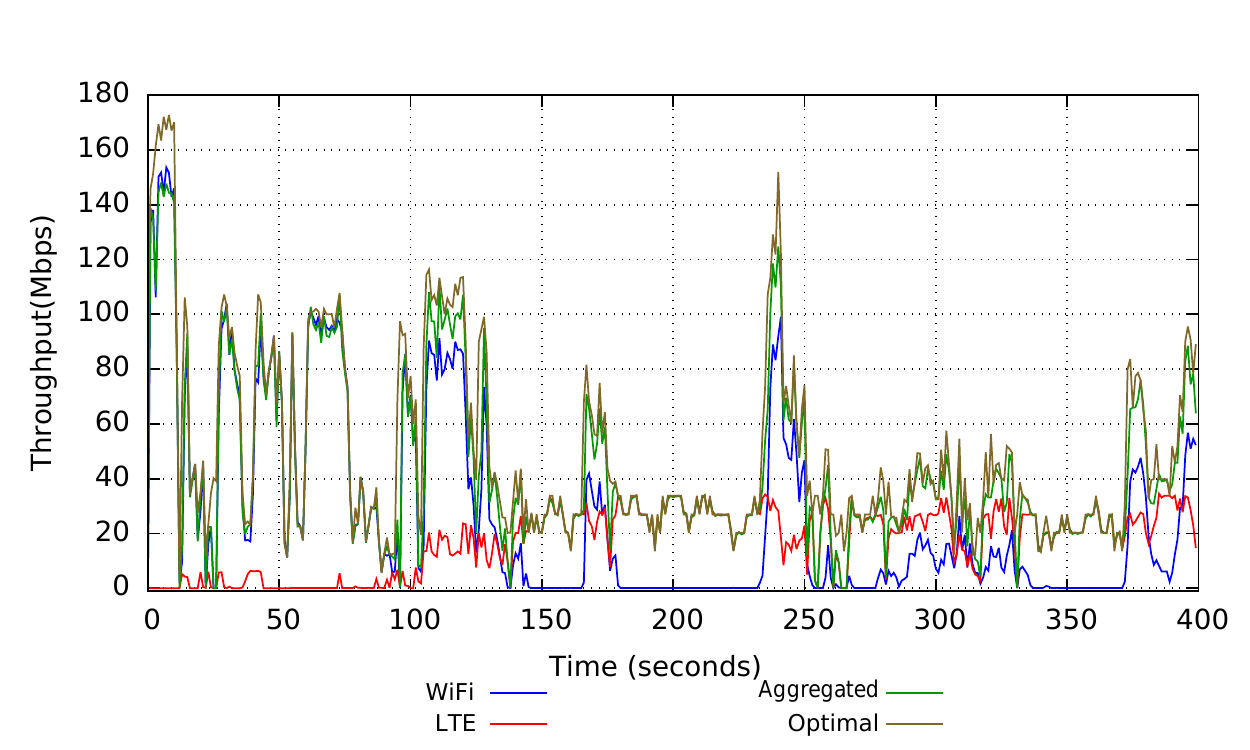}
  \caption{Throughput of the proposed system using WRR\\(Realistic dynamic capacity scenario)}
  \label{fig:UQ-WRRTrafficFlow_After}
  \end{figure}

To validate the impact of socket timeout on the measured bandwidth, the WRR load balancing algorithm was re-run in our proposed system. Figure~\ref{fig:UQ-WRRTrafficFlow_After} shows the throughput obtained by applying the WRR algorithm on the proposed system with the application socket timeout of 10 seconds. Compared to the previous WRR experiment in Figure~\ref{fig:UQ-WRRTrafficFlow}, the aggregated throughput is almost near to the optimal one.

\begin{figure}[t]
  \centering
   \captionsetup{justification=centering}
  \includegraphics[ width=3.35in]{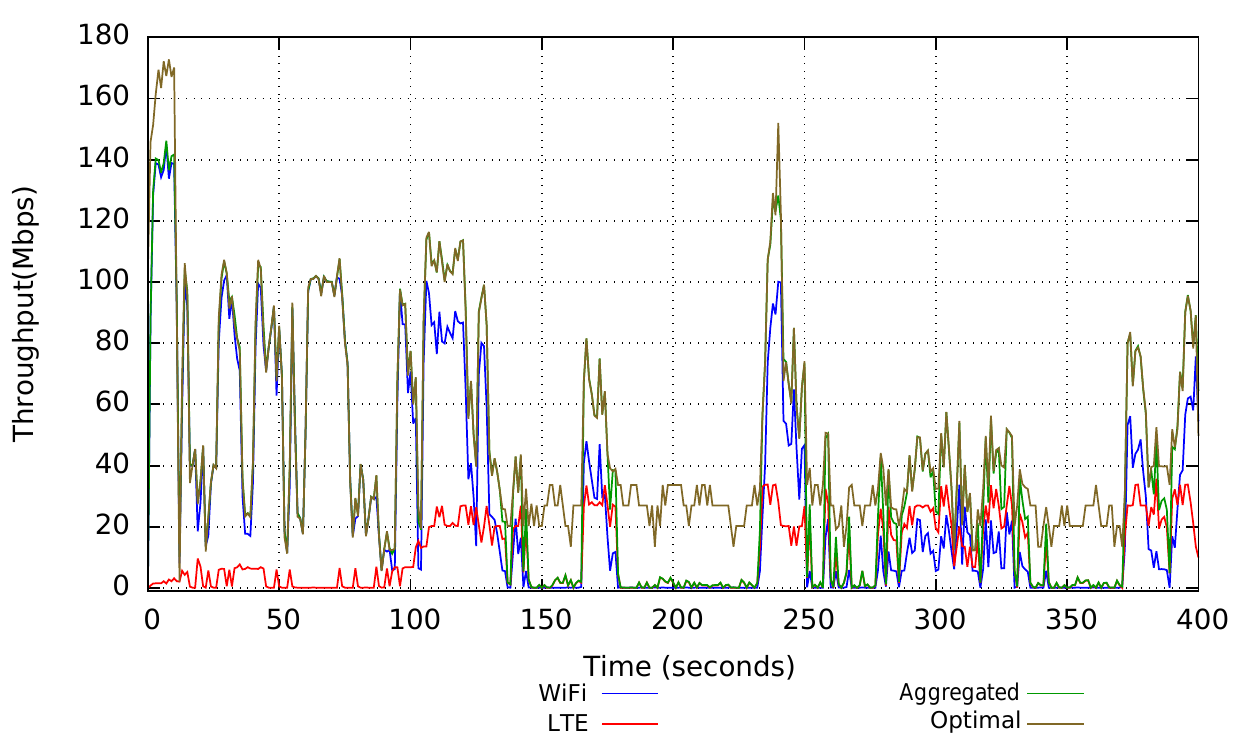}
  \caption{Throughput of the last hop links using MPTCP\\(Realistic dynamic capacity scenario)}
  \label{fig:UQ-MPTCPTrafficFlow}
\end{figure}

After evaluating the proposed system with the WRR algorithm, we would like to compare our approach with MPTCP running over the topology proposed in Figure~\ref{fig:Network-Topo}. Figure~\ref{fig:UQ-MPTCPTrafficFlow} shows the throughput of applying MPTCP over 400 seconds. As it is seen, the Aggregated throughput collected from WiFi and LTE links drops when the WiFi throughput plummeted to zero regardless of the other links throughput. This could be due to the drop in the primary link throughput, which is WiFi in our implementation since the MPTCP uses a decoupled congestion control algorithm (as explained earlier in section \ref{subsec:Simple Dynamic capacity(Step Function)}) that affects the overall throughput.

\begin{figure}[t]
  \centering
  \captionsetup{justification=centering}
  \includegraphics[width=3.35in]{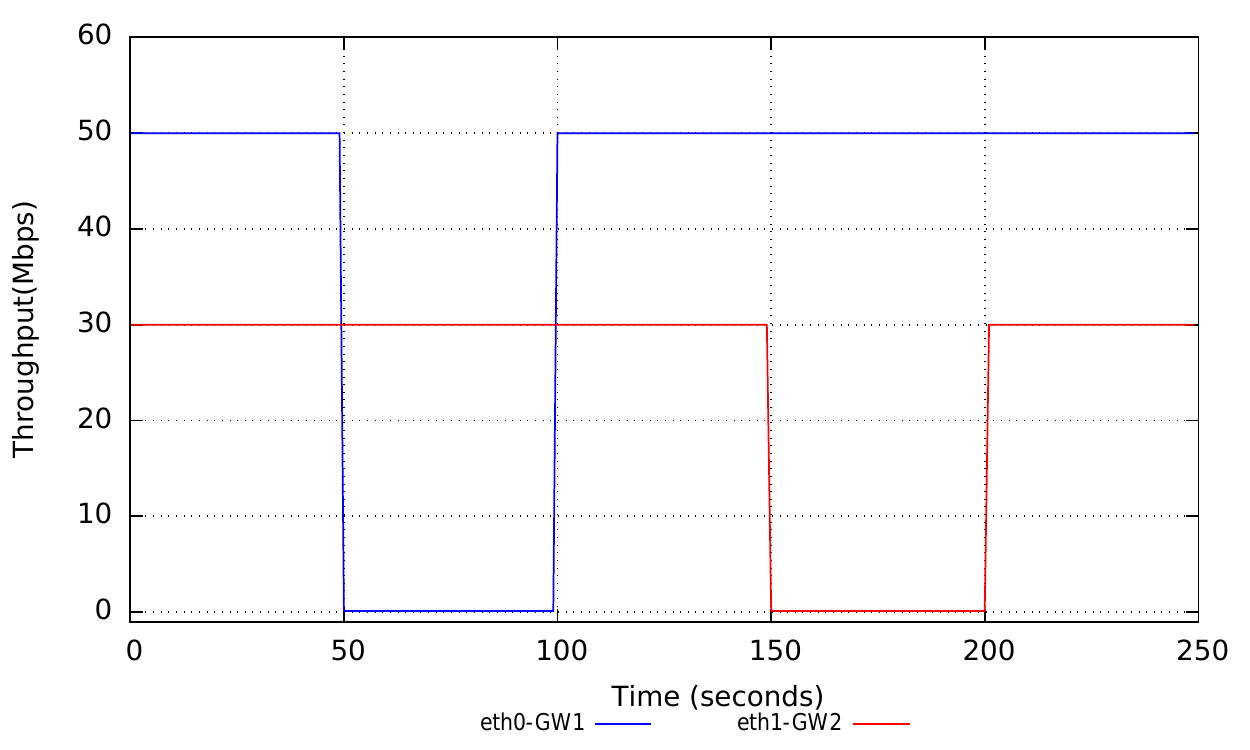}
  \caption{MPTCP Validation Synthetic Bandwidth}
  \label{fig:MPTCP Validation BW}
\end{figure}

\begin{figure}[t]
  \centering
   \captionsetup{justification=centering}
  \includegraphics[width=3.35in]{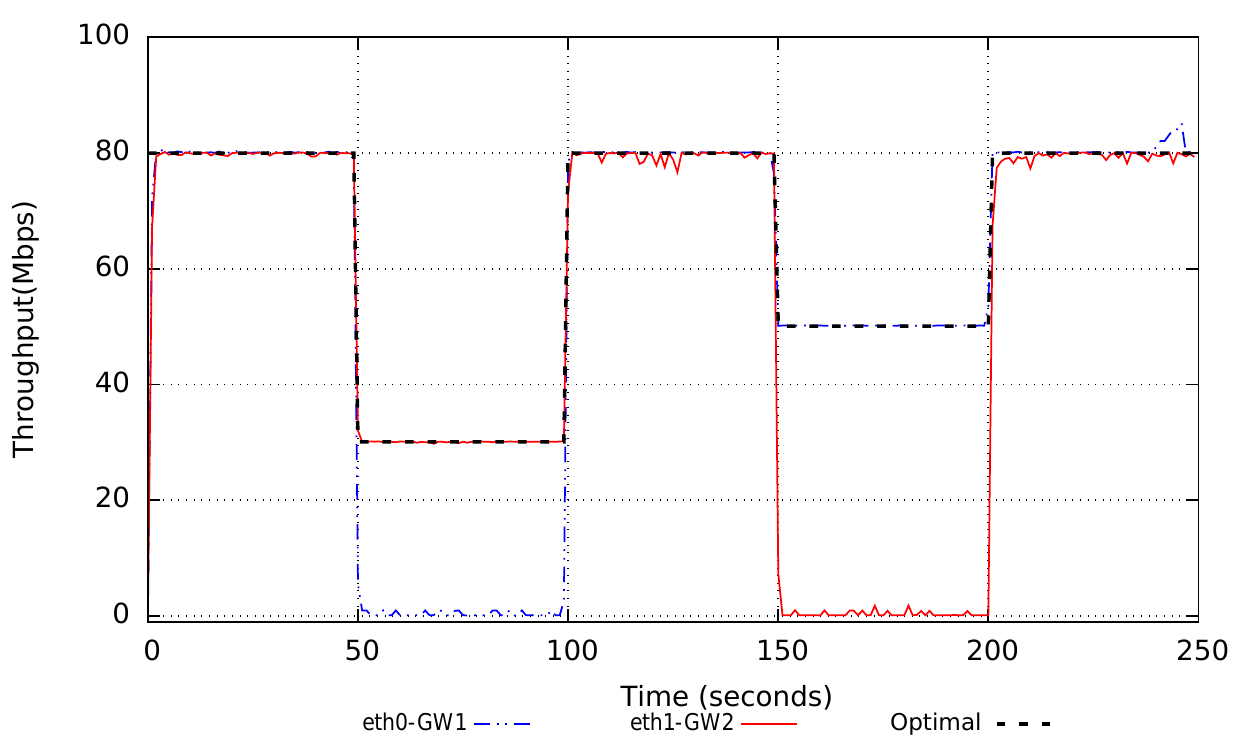}
  \caption{MPTCP throughput validation with different\\ Primary link selections}
  \label{fig:MPTCP-Eth0-Eth1-Defaults}
\end{figure}

To prove this hypothesis, we run our experiment by alternating the primary link in MPTCP as in the previously stated topology. Multiple MPTCP connections were established on the end-host and the server. The MPTCP was configured on the end-host and the server such that paths using (eth0-GW1 and eth1-GW2) links can be alternatively selected. We also chose a steady bandwidth scenario for 250 seconds as depicted in Figure~\ref{fig:MPTCP Validation BW}. The bandwidth drops to zero for some time to show the impact of primary link bandwidth over MPTCP performance. As in previous experiments, the Apache benchmark $ab$ tool is used to generate the traffic.

Figure~\ref{fig:MPTCP-Eth0-Eth1-Defaults} shows the MPTCP throughput performance for different primary links and it is compared with the optimal throughput that should be satisfied. In the case of eth0-GW1 primary link, the aggregated throughput is accumulative of both links throughput unless when the bandwidth of End-host--GW1 drops to zero (from 50 to 100 seconds). In that interval, the throughput fluctuates around zero although bandwidth of eth1-GW2 link at that interval is 30Mbps. Nonetheless, dropping the bandwidth of eth1-GW2 link (150-200 seconds) does not affect the performance of MPTCP. The same trend can be observed when the primary interface is flipped to eth1-GW2. We obtained a total throughput of both links except between 150-200 seconds interval where it is zero due to a drop in eth1-GW2 bandwidth. Indeed, plummeting the bandwidth of the primary interface severely impact on multi-link MPTCP performance.

\begin{figure}[t]
  \captionsetup{justification=centering}
  \centering
  \includegraphics[width=3.35in]{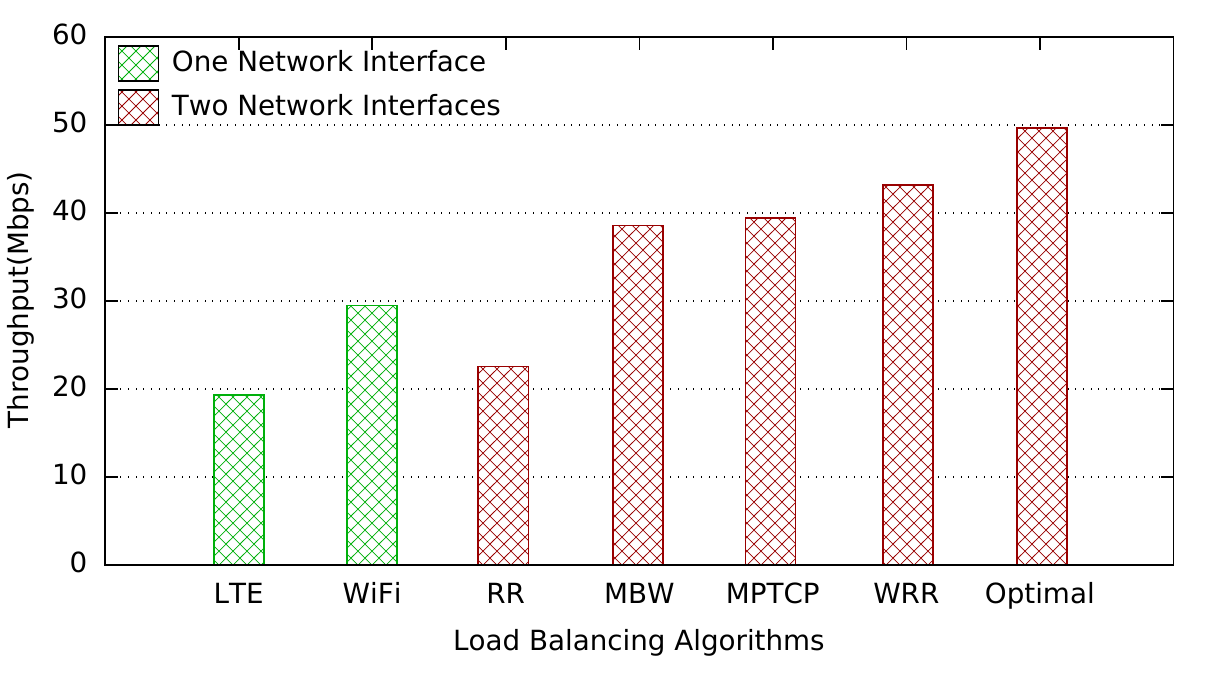}
  \caption{Throughput of different traffic distribution schemes\\(Realistic dynamic capacity scenario)}
  \label{fig:UQ-LB-Algorithms}
\end{figure}

To sum up, we evaluate the performance of different load balancing in the above set up. Figure~\ref{fig:UQ-LB-Algorithms} shows the throughput obtained by applying various load balancing algorithms and compared with MPTCP and the optimal throughput. As noticed, the throughput of utilising one network link is near to than 20Mbps and 30Mbps for LTE and WiFi cases, respectively. However, with multiple network interfaces, the throughput is higher except when RR is used. The RR achieved 22.5 Mbps slightly higher than the LTE link due to unequal traffic distribution between links. For MBW, the achieved throughput is $\approx$ 38.5Mbps, higher than RR and closer to MPTCP (39Mbps). Finally with WRR $\approx$ 43.2Mbps, the achieved throughput is even higher than the previous algorithms e.g. 55\% and 10\% more than single network interface and  MPTCP, respectively. Note that, this is the closest throughput to the optimal case in which both links bandwidth are simply added.

To conclude, the results show a significant potential of leveraging SDN based end-host for load balancing. From the extensive experimental evaluation, our findings clearly show that the SDN based load management over end-hosts outperforms several other load balancing approaches including MPTCP, in different link capacity scenarios.

\section{Conclusions}
\label{sec:Conclusion}

In this paper, we have explored the potential of controlling multi-homed devices for achieving load balancing via utilising SDN concept. Compared with other related works, the proposed approach is completely transparent to the end-host applications, the service provider as well as the server-side. Furthermore, it can be deployed by simply installing and configuring a software OpenFlow switch (e.g. OVS) and a (lightweight) SDN controller on an end-host, which includes a wide range of devices (PCs, smartphones, tablets, etc.) This approach has significant benefits in many potential applications (web browsing, file download, etc.) by providing a greater aggregate network throughput and increased reliability, resulting in an improved quality of user experience.

The proposed system has been evaluated by various load balancing algorithms conducted with different link capacity configurations. The evaluation results meet all of the expected goals and most importantly achieves a throughput improvement of more than (55\% compared to the only single network interface and 10\% compared to MPTCP). 
%


\end{document}